\newcommand{\average}[1]{\langle{#1}\rangle}
\newcommand{\Sn}[1]{S^{(#1)}}
\newcommand{\Hn}[1]{H^{(#1)}}
\newcommand{\kn}[1]{k^{(#1)}}
\newcommand{\nn}[1]{n^{(#1)}}
\newcommand{\half}{\nicefrac{1}{2}}
\newcommand{\ts}[1]{\tilde{\sigma}_{#1}}
\newcommand{\diff}[2]{\frac{d#1}{d#2}}

\newcommand{\diffst}[1]{\diff{\tilde\sigma_{#1}}{t}}
\newcommand{\Iss}{\average{I}_{\text{ss}}}
\newcommand{\Jp}{J_p}
\newcommand{\ca}{\cos\alpha}
\newcommand{\sa}{\sin\alpha}
\newcommand{\can}[1]{\cos^{#1}\alpha}
\newcommand{\san}[1]{\sin^{#1}\alpha}
\newcommand{\steady}[2][]{\average{\rho^{#1}_{#2}}_{\text{ss}}}
\newcommand{\nba}{\bar{n}_{BA}}
\newcommand{\gba}{\gamma_{BA}}
\newcommand{\mat}[1]{\uuline{#1}}

\renewcommand{\vector}[1]{{\boldsymbol{#1} }}

\documentclass[a4paper,prb,amsmath,amssymb, superscriptaddress, showpacs, twocolumn]{revtex4}

\usepackage{graphicx, braket, mathrsfs, nicefrac, epsfig, multirow, ulem, verbatim}

\DeclareMathOperator{\sgn}{sgn}
\DeclareMathOperator{\Tr}{Tr}

\begin{document}

\title{Charge noise at Cooper-pair resonances}
\author{P.~G.  Kirton}
\affiliation{School of Physics and Astronomy, University of Nottingham, Nottingham NG7 2RD, U.K.}
\author{M. Houzet}\affiliation{CEA, INAC, SPSMS, F-38054 Grenoble, France}
 \author{F. Pistolesi} \affiliation{CPMOH, Universit\'{e} de Bordeaux I \& CNRS,  F-33405 Talence, France}
 \affiliation{Laboratoire de Physique et Mod\'{e}lisation des Milieux Condens\'{e}s \\
Universit\'{e} Joseph Fourier \& CNRS,  F-38054 Grenoble, France}

\author{A.~D. Armour}  \affiliation{School of Physics and Astronomy, University of Nottingham, Nottingham NG7 2RD, U.K.}
\date{\today}

\begin{abstract}
 We analyze the charge dynamics of a superconducting single-electron transistor (SSET) in the regime where charge transport occurs via Cooper-pair resonances. Using an approximate description of the system Hamiltonian, in terms of a series of resonant doublets, we derive a Born-Markov master equation describing the dynamics of the SSET. The average current displays sharp peaks at the Cooper-pair resonances and we find that the charge noise spectrum has a characteristic structure which consists of a series of asymmetric triplets of peaks. The strongest feature in the charge noise spectrum is the triplet of peaks centered at zero frequency which has a peak spacing equal to the level separation within the doublets and is similar to the triplet in the spectrum of a driven, damped, two-level system. We also explore the back-action that the SSET charge noise would have on an oscillator coupled to the island charge, measurement of which provides a way of probing the charge noise spectrum.
\end{abstract}


\maketitle

\section{Introduction}

Mesoscopic superconducting circuits in which there is an interplay between Josephson tunneling and electrostatic charging effects display a wide range of interesting behaviors. The quantum coherence and non-linearity generated by Josephson junctions make superconducting circuits obvious candidates for qubits.\cite{makhlin:01,korotkov:09} However, the usefulness of qubits depends strongly on the extent to which the level of dissipation and the attendant decoherence can be minimized.\cite{korotkov:09} In contrast, there are many other applications of superconducting circuits, as measuring devices\cite{korotkov:96,clerk:02,lu:03,sillanpaa:04,lahaye:04} or as coolers,\cite{naik:06} where irreversible transport is necessary and hence dissipation plays an essential positive role.

The superconducting single electron transistor\cite{averin:91} (SSET) is an example of a superconducting device with applications where both coherence and dissipation are important. The SSET consists of a small superconducting island linked to superconducting leads by Josephson junctions. A voltage gate coupled to the island can be used to control the flow of charge. The strong dependence of the charge transport through the SSET island on the properties of the gate makes it an ideal sensing device; it can act as either an electrometer\cite{korotkov:96,clerk:02,lu:03,sillanpaa:04} or, (when the gate is mechanically compliant) as a displacement detector.\cite{lahaye:04,naik:06} Sensing typically involves a coupling between the SSET island charge and the degree of freedom being measured, hence the charge noise spectrum\cite{clerk:02,clerk:05,clerk:08}  determines the back-action of the SSET on the measured system (within the linear response regime). Although the back-action is a nuisance in the context of measurement, it can be used to manipulate the state of the measured system:\cite{clerk:05,blencowe:05,rodrigues:07} one recent experiment used the back-action from a SSET to cool a nanomechanical resonator,\cite{naik:06} whilst another demonstrated the production of laser-like states of self-sustained oscillation in an electrical resonator.\cite{astafiev:07} From another perspective the back-action of the SSET can be seen as providing an efficient means of probing its quantum noise properties.\cite{clerk:08,xue:09}

The SSET supports a wide range of different current carrying processes depending on the choice of operating
point.\cite{aleshkin:90,maassen:91,maassen:91a,haviland:94,joyez:94,siewert:96,fitzgerald:98,hadley:98} When a large voltage is applied
($V\ge 4\Delta/e$, where $\Delta$ is the superconducting gap), it becomes energetically favorable to break up the Cooper-pairs and the transport is 
dominated by inelastic (incoherent) tunneling of quasiparticles at both junctions. At lower voltages the transport involves Josephson tunneling of 
Cooper-pairs, though dissipation is still required to generate a dc current. In the regime where $V \sim \Delta/e$, current resonances known as 
the Josephson quasiparticle\cite{aleshkin:90,maassen:91,hadley:98} (JQP) and double Josephson quasiparticle\cite{hadley:98,clerk:02} (DJQP) cycles 
combine coherent Cooper-pair tunneling and dissipative quasiparticle tunneling. The JQP and DJQP cycles have attracted much recent attention as the 
resonant and (partly) coherent nature of the transport leads to measurement sensitivities approaching the quantum limit, as well as a range of 
interesting back-action effects.\cite{clerk:02,lahaye:04,clerk:05,blencowe:05,naik:06,bennett:06,astafiev:07,rodrigues:07,xue:09,choi:03}

In this article we focus instead on Cooper-pair resonances\cite{maassen:91,maassen:91a,haviland:94,joyez:95,siewert:96,billangeon:07,leppakangas:08}
which occur at even lower voltages than the JQP and DJQP cycles and involve the coherent transfer of one or more Cooper-pairs across the two SSET junctions. At these lower voltages quasiparticles are almost completely absent and dissipation and decoherence are dominated by the electromagnetic environment of the SSET. Cooper-pair resonances are known to give rise to sharp features in the current voltage characteristics of the SSET,\cite{maassen:91,maassen:91a,haviland:94,joyez:95,siewert:96,billangeon:07,leppakangas:08} though much less is known about their quantum noise properties.

We develop a simple model for the SSET system, valid at low voltages and for embedding circuits with resistances much less than the resistance
quantum, $R_Q=h/4e^2$. This model, which has important similarities with the standard description of resonance fluorescence of quantum
optics, divides the energy levels of the system into pairs of resonant levels with the spacing within a doublet much less than the spacing between doublets.\cite{cohen:92,joyez:95} We derive the master equation of the SSET within the Born-Markov approximations and then  calculate both the average current and the charge noise spectrum in the vicinity of the Cooper-pair resonances. The dominant feature in the charge noise spectrum is a triplet of peaks centered at zero frequency that is a characteristic of a driven two-level system,\cite{jaehne:08, wang:09} additional weaker triplets of peaks also occur at much higher frequencies. We also investigate the back-action that the Cooper-pair resonances would have on an oscillator coupled to the SSET island. The asymmetry in the triplet of peaks centered on zero frequency means that the resonances could be used to cool a mechanical resonator with frequency in the MHz range though not all the way down to its ground state.

The organization of this paper is as follows. In Sec.\ \ref{sec:model} we describe our model of the SSET at low voltages, discussing the origin of the Cooper-pair resonances and the way in which the electromagnetic environment couples to the charge passing through the transistor. Then in Sec.\ \ref{sec:ME} we show how the Hamiltonian of the SSET close to a Cooper-pair resonance can be transformed to give a description in terms of a set of doublet states. We then use standard approximations to derive a master equation that describes the SSET charge variables in the presence of dissipation due to the electromagnetic environment. Next, in Sec.\ \ref{sec:current}, we use the master equation to derive the average current at the resonances. The charge noise spectrum of the SSET is calculated in Sec.\ \ref{sec:CN} and the back-action is discussed. Finally we give our conclusions in Sec.\ \ref{sec:conclusions}. The appendix contains additional details about certain aspects of the calculations.

\section{Model System}
\label{sec:model}

\begin{figure}
 \center{\epsfig{file=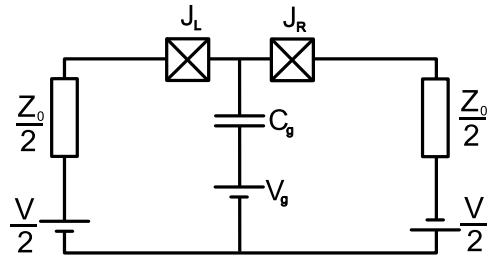, width=0.35\textwidth}}
 \caption{Circuit diagram of the SSET. The SSET island is linked to the leads by the Josephson junctions, $J_{L(R)}$, and is coupled to the voltage gate by the capacitance $C_g$. The bias voltage, $V$, and impedances in the circuit, $Z_0$, are taken to be distributed symmetrically. }\label{fig:Circuitdiag}
\end{figure}

The SSET is shown schematically in Fig.\ \ref{fig:Circuitdiag}. For simplicity, we assume that the drain-source bias, $V$, is applied symmetrically and take the junctions to have equal capacitances, $C_J$, and Josephson energies, $E_J$. A voltage, $V_g$, is applied to the gate which has capacitance,  $C_g$, (assumed to be much less than $C_J$).  The electromagnetic environment of the SSET is modeled by the impedance $Z_0$. We assume that the  charging energy of the island, $E_C=4e^2/2C_{\Sigma}$, with $C_{\Sigma}=2C_J+C_g$, is the dominant energy scale in the system, so that $E_C\gg k_{\rm B}T, E_J$, where  $T$ is the temperature.  In this limit the SSET is best described in terms of charge states, using the two quantum numbers, $n=N_L-N_R$, the excess  number of Cooper-pairs on the island, and $k=(N_L+N_R)/2$, the average number of Cooper-pairs which have traveled through the system,  with $N_{L(R)}$ the number of Cooper-pairs which have passed through the left(right)-hand junction of  the SSET.\cite{maassen:91a, joyez:95, siewert:96} The large charging energy means that only the two states with values of $n$ closest to the gate induced polarization charge, $n_g=C_gV_g/2e$, have an appreciable chance of being occupied.

The full Hamiltonian of the system can be written as,
\begin{equation}
 H=H_S + H_{\text{int}} + H_{\text{env}},
\end{equation}
where $H_{\rm env}$ is the Hamiltonian of the electromagnetic environment and the SSET Hamiltonian, $H_S$, consists of two parts $H_S=H_{\text{ch}}+H_J$. The charging Hamiltonian of the island is
\begin{equation}
H_{\text{ch}}=\sum_{n=0,1}\sum_{k}[E_C(n-n_g)^2-2eVk]\ket{n,k}\bra{n,k}, \label{eqn:Hch}
\end{equation}
where we have taken $0\leq n_g<1$, so only the states $\ket{0,k}$ and $\ket{1,k}$ need to be considered.

The states $\{\ket{n,k}\}$ separate into two ladders: $\{\ket{0,k},\ket{1,k+\half}\}$ and $\{\ket{0,k+\half},\ket{1,k}\}$,
where $k$ is now an integer. Josephson coupling links together adjacent members of the same ladder, but does not connect states from different ladders.
Assuming that quasiparticle tunneling (which can link the ladders) is negligible, we need only consider one of the sets of states. Choosing the ladder
$\{\ket{0,k},\ket{1,k+\half}\}$, the Josephson coupling between states is given by,\cite{joyez:95,leppakangas:08}
\begin{equation}
H_J=-J\sum_{k}(\ket{0,k}\bra{1, k+\nicefrac{1}{2}} + \ket{0,k}\bra{1, k-\nicefrac{1}{2}} + h.c.), \label{eqn:HJ}
\end{equation}
where $J=E_J/2$.

We assume that the dominant source of dissipation and decoherence of the SSET is the electromagnetic environment, modeled by the impedances in the leads connecting it to the voltage sources. The impedances lead to fluctuations in the drain-source voltage, $\delta V$, which couple to the system operator, $k$, and give rise to  the interaction Hamiltonian,
\begin{equation}
  H_{\text{int}}=-2ek\delta V.
\end{equation}
The effects of $\delta V$ on the SSET are determined by the spectrum of the voltage fluctuations which takes the form,\cite{ingold:92, joyez:95, makhlin:03, leppakangas:08, nazarov:09}
\begin{align}
S_V(\omega)&=4e^2\int_{-\infty}^\infty \average{\delta V(t)\delta V(0)}{\rm e}^{i\omega t} \;dt,\nonumber\\
&=\frac{8e^2  \hbar\omega }{1-{\rm e}^{-\hbar\omega/k_{\rm B}T}}{\rm Re}[Z_T], \label{eqn:specdensity}
\end{align}
where $Z_T=(Z_0^{-1}-i\omega C_J^2/C_\Sigma)^{-1}$ is the total effective impedance seen at the junctions. At the low frequencies which turn out to be relevant for the system dynamics, $\omega\ll ({\rm Re}[Z_T]C_J)^{-1}$, we can take $\rm{Re}[Z_T]= \rm{Re}[Z_0]$. We further assume that the embedding circuit provides a low, real (Ohmic) impedance, $Z_0=R\ll R_Q$, such as would be generated by a transmission line.\cite{clerk:08} Our description can easily be extended to take into account a finite impedance in series with the gate voltage,\cite{leppakangas:08} but since we take the limit $C_g\ll C_J$ this has a much weaker influence and so we neglect it here.

\begin{figure}
 \center{ \epsfig{file=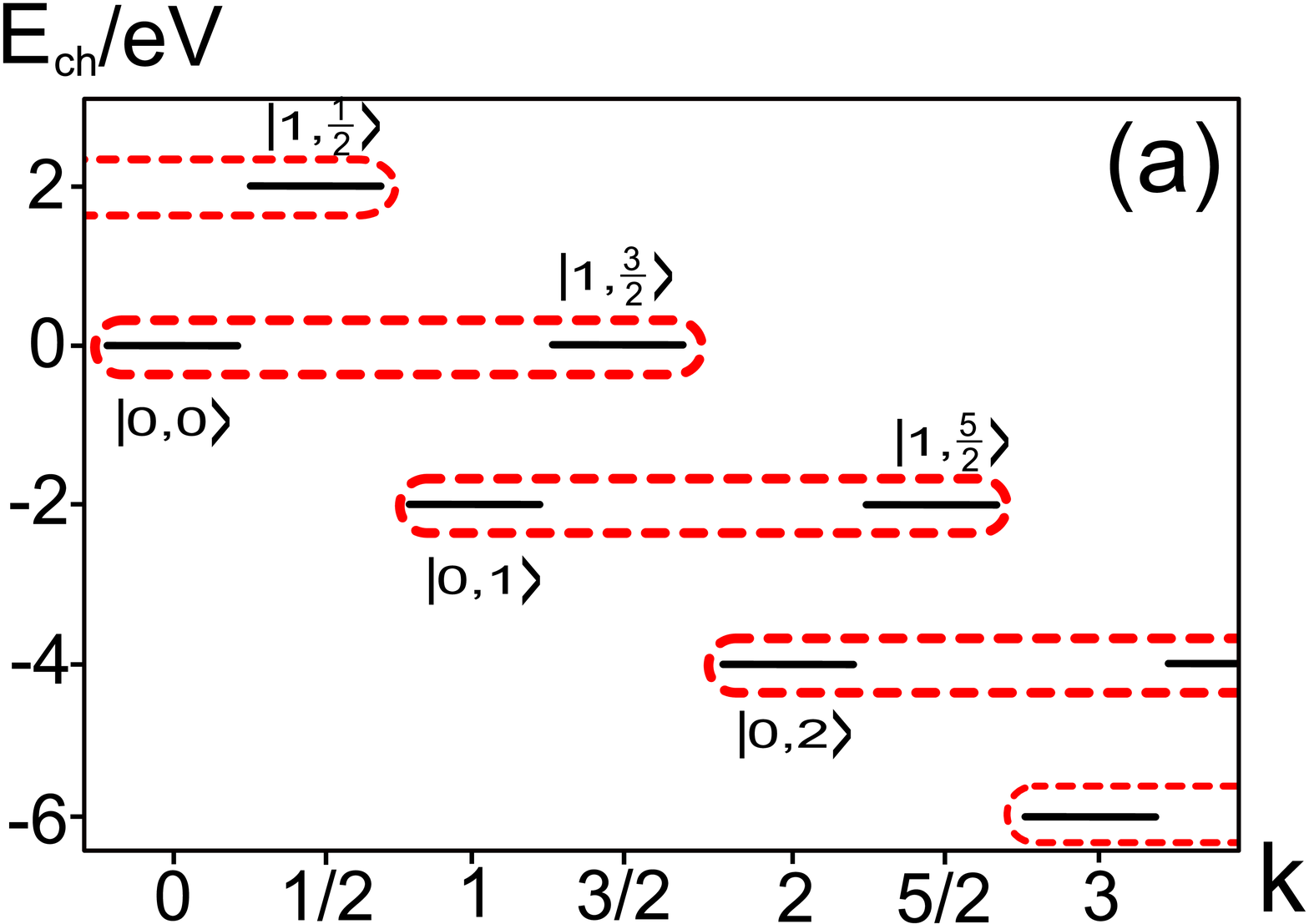, width=0.23\textwidth} \epsfig{file=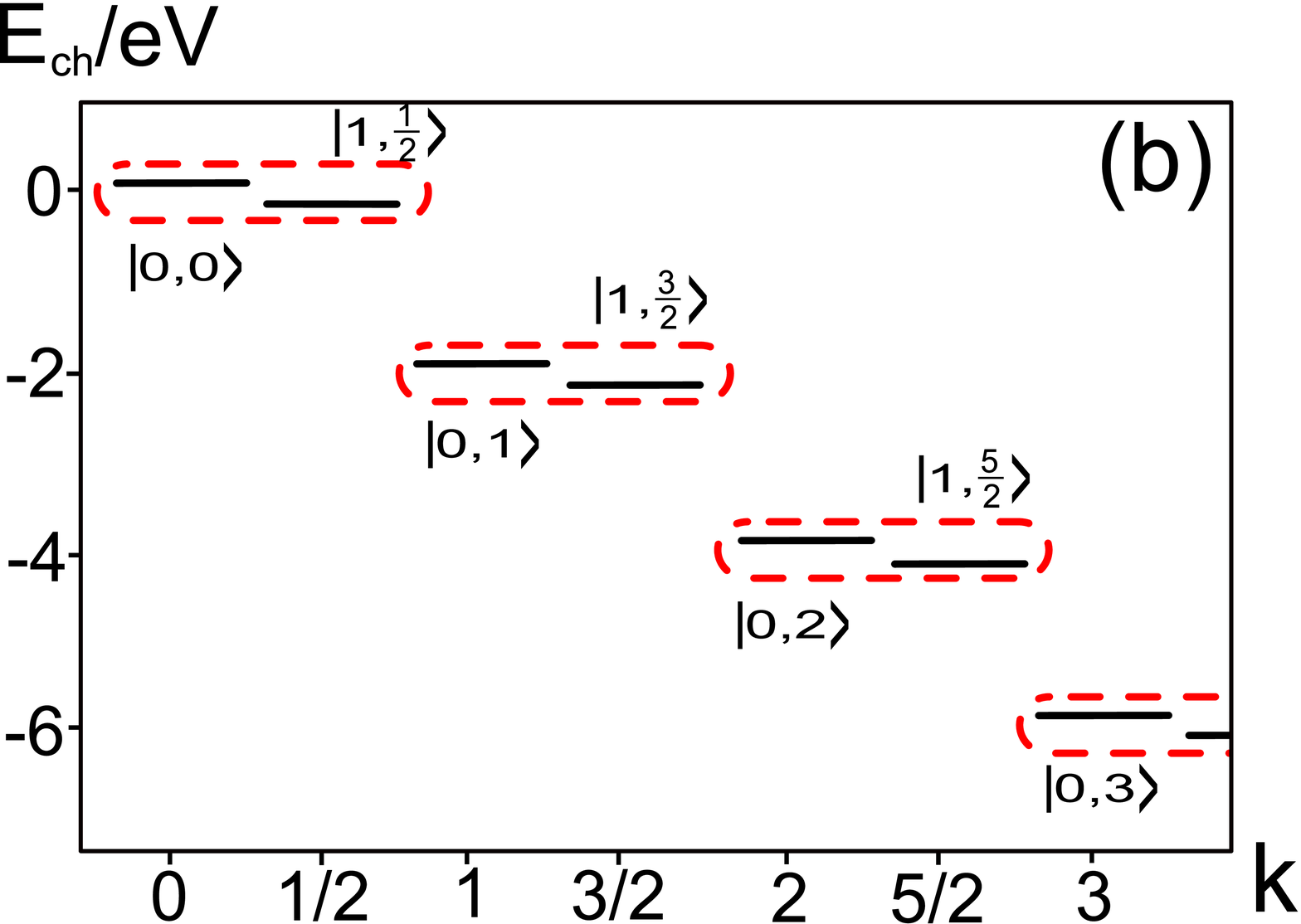 ,width=0.23\textwidth}}
 \caption{(Color online) Energy levels of $H_{\rm ch}$, (a) with $V=V^{(1)}_{\rm res}$ at the $p=1$ resonance and (b) with $V\gtrsim V^{(0)}_{\rm res}$ for $p=0$. 
 The dashed lines enclose the doublets: the energy levels which are almost degenerate near the given resonance.
 }\label{fig:presonances}
\end{figure}

The voltage dependence of the charging energy leads to resonances where the eigenvalues of the charging Hamiltonian, $H_{\text{ch}}$ [Eq.\ \ref{eqn:Hch}], become
degenerate. The charging energies of the states $\ket{0,k}$ and $\ket{1,k+p+\half}$, where $p=0,1,2,\hdots$,
become degenerate at particular values of the drain-source voltage, $V=V_{\rm res}^{(p)}$, given by
$(2p+1)eV_{\rm res}^{(p)}=E_C(1-2n_g)$. The energy levels near the $p=0$ and $p=1$ resonances are illustrated in Fig.\ \ref{fig:presonances}.

Close to degeneracies in the charging energy the Josephson coupling becomes important and the pairs of states $\ket{0,k}$ and $\ket{1,k+p+\half}$ become strongly mixed forming doublets. The interaction with the environment can then cause the system to decay into the neighboring doublets with lower energy. Taken together the coherent evolution
and decay form a cascade in which $k$ increases systematically and hence a dc current flows. Away from the resonances the coherent evolution is 
suppressed and decay processes cannot take the system to ever higher $k$ values so the current is also suppressed. Thus, the degeneracies in the charging energy lead to resonances in the current.\cite{maassen:91, haviland:94, joyez:95, toppari:07, leppakangas:08} The one exception to this picture arises for the $p=0$ resonance where one can see from Fig.\  \ref{fig:presonances}(b) that for voltages above resonance, $V>E_C|1-2n_g|/e$, the system can move indefinitely to larger values of $k$ via incoherent decay processes alone and hence in this case the resonance becomes strongly broadened on one side.\cite{joyez:95,leppakangas:08}

\section{Master equations}
\label{sec:ME}

Having seen how and where the Cooper-pair resonances arise, we now proceed to obtain a detailed quantitative description of the charge dynamics of the SSET that includes the dissipation and decoherence induced by the electromagnetic environment. As a first step we use a unitary transformation method to derive an effective Hamiltonian which provides a systematic way of accounting for the coherent effect of the Josephson coupling between resonant states. We then proceed to derive the master equation for the SSET tracing out the environment using the Born-Markov approximations. The resulting master equation can then be used to derive a much simpler equation that describes just the SSET island charge.

\subsection{Effective Hamiltonian}

Close to the $p$-th Cooper-pair resonance, and provided that the voltage, $V$, is not too small, the eigenvalues of the SSET charging Hamiltonian are grouped into doublets,
 $\{\ket{0,k},\ket{1,k+p+\half}\}$, with the spacing between members of a given doublet much less than the spacing between the doublets. The main effect of the Josephson Hamiltonian will be to introduce couplings between states within
each doublet. Since the full system Hamiltonian cannot be diagonalized exactly, we treat the Josephson coupling as a perturbation and use a unitary transformation to derive an effective Hamiltonian which takes into account the mixing it induces between states within a
doublet.\cite{cohen:92}

We seek a unitary transformation, $U$, such that, $H_S'=UH_SU^{\dagger}$ is block diagonal in the space of the doublets. This transformation is found as a perturbation series in $J$, and we keep only the leading order contributions   (details are given in the appendix). This results in an effective system Hamiltonian which is block-diagonal in the pairs of nearly-resonant states $\{\ket{0,k},\ket{1,k+p+\half}\}$. Each block takes the form,
\begin{equation}
   H'_{k}=  \begin{pmatrix}
     \bar{E} - \Delta E & \Jp  \\
      \Jp & \bar{E} +\Delta E
     \end{pmatrix}, \label{eqn:defineH'}
 \end{equation}
where $\bar{E}=-2eVk$ is the average charging energy of the resonant states, and $\Jp$ is the high order coupling between the states. For $p\geq1$,
\begin{equation}
\label{eqn:jp}
 \Jp=\left(-1\right)^p\frac{J^q}{\left(2eV_{\rm res}^{(p)}\right)^{2p}\left(p!\right)^2}
\end{equation}
where $q=2p+1$. For the case $p=0$, $\Jp=J$.  The splitting between resonant states, $\Delta E$, is given by,
\begin{equation}
 \Delta E=\frac{E_C(1-2n_g)-qeV}{2}+\frac{J^2}{eV_{\rm res}^{(p)}}\frac{2q}{q^2-1}.
\end{equation}
The first term is the electrostatic energy difference between the states and the second term is a correction which
arises at second order in the perturbation calculation\cite{footnote3}. For $p=0$ the second order correction is given by $J^2/(eV_{\rm res}^{(0)})$.

The block Hamiltonians are diagonalized by a rotation, $U_{\alpha}={\rm e}^{-i\sigma_y \alpha}$, where $\sigma_y$ is the usual Pauli matrix, to give the eigenstates of the doublets,
\begin{subequations}
\begin{gather}
  \ket{a,k}=\cos \alpha \ket{0,k} + \sin \alpha \ket{1, k+\nicefrac{(2p+1)}{2}}, \\
  \ket{b,k}=-\sin{\alpha} \ket{0,k} +\cos \alpha \ket{1, k+\nicefrac{(2p+1)}{2}},
 \end{gather}
 \label{eqn:eigenstates}
\end{subequations}
and the corresponding eigenenergies,  $E_{a,k}=\bar{E}-\Delta E'$, $E_{b,k}=\bar{E} + \Delta E'$,
where $\alpha$ is defined by
\begin{subequations}
\begin{gather}
 \sin 2\alpha=\frac{-J_p}{\Delta E'}, \\ \cos 2\alpha=\frac{\Delta E}{\Delta E'},
 \end{gather}
 \label{eqn:sincosalpha}
\end{subequations}
and the energy level splitting is
 \begin{equation}
   \Delta E'=\sgn(\Delta E)\sqrt{\Delta E^2+J_p^2}, \label{eqn:DE'}
 \end{equation}
which changes sign at the resonance.

Note that the description of the system in terms of doublets is only valid within a region around each resonance. The spacing between energy levels in the doublet should be much smaller than the spacing between the doublets, $|\Delta E'|\ll eV$.

\subsection{Born-Markov description}

We now use the block-diagonal form of the Hamiltonian to derive the master equation for the SSET. We assume that the interaction between the SSET and the bath is weak,\cite{makhlin:03} $R\ll R_Q$, and that the bath has a sufficiently dense spectrum of levels that the standard Born and Markov approximations can be made.\cite{cohen:92}

Written in terms of the eigenstates of the system Hamiltonian, the Born-Markov master equation for the components of the SSET density operator, $\sigma$, takes the form,\cite{cohen:92}
\begin{equation}
\diff{\tilde{\sigma}_{\mu\nu}}{t}=\sum_{\mu'\nu'}^{\rm sec}\mathscr{R}_{\mu\nu\mu'\nu'}\tilde{\sigma}_{\mu'\nu'},\label{eqn:MEdef}
\end{equation}
where the tilde denotes the interaction picture and the sum is over only the secular\cite{footnote} terms for which $\omega_{\mu\nu}=\omega_{\mu'\nu'}$, with $\omega_{\mu\nu}$ the frequency difference between eigenstates $\mu$ and $\nu$.  The coupling tensor $\mathscr{R}_{\mu\nu\mu'\nu'}$ is given by
\begin{widetext}
\begin{multline}
\mathscr{R}_{\mu\nu\mu'\nu'}=-\int_0^\infty d\tau  \left[g(\tau)\left(\delta_{\nu\nu'}\left[\sum_n k_{\mu n}'k_{n\mu'}'{\rm e}^{i\omega_{\mu'n}\tau}\right]
-k_{\mu\mu'}'k_{\nu'\nu}'{\rm e}^{i\omega_{\mu'\mu}\tau}\right)\right. \\ \left.+ g(-\tau) \left(\delta_{\mu\mu'}\left[\sum_n k_{\nu'n}'k_{n\nu}'{\rm e}^{i\omega_{n\nu'}\tau}\right]
-k_{\mu\mu'}'k_{\nu'\nu}'{\rm e}^{i\omega_{\nu\nu'}\tau}\right)\right], \label{eqn:Rabcd}
\end{multline}
\end{widetext}
where $k'=UkU^{\dagger}$ and $g(\tau)=4(e/\hbar)^2\average{ \delta V(\tau)\delta V(0)}$ is the correlation function of the electromagnetic environment whose properties were specified in Eq. \eqref{eqn:specdensity}.

The transformed operator $k'$ (given explicitly in the appendix) takes the form of a power series in $J$,
\begin{equation}
k'=\kn{0}+\kn{1}+\kn{2}+\ldots,
\end{equation}
and we proceed by expanding the terms of the form $k'_{\mu\mu'}k'_{\nu\nu'}$ in $\mathscr{R}$ up to second order in $J$. The zeroth order term $\kn{0}_{\mu\mu'}\kn{0}_{\nu\nu'}$ is diagonal in the charge state basis and generates dephasing of the charge states. For states within the same doublet, this leads to dissipative transitions between the eigenstates (intra-doublet transitions). The next non-zero contribution comes from terms of the form $\kn{1}_{\mu\mu'}\kn{1}_{\nu\nu'}$, which link states from a given doublet with states in the nearest neighbor and next-nearest neighbor doublets, leading to inter-doublet transitions. The same generic description applies to all the resonances with $p\geq 1$, but for the $p=0$ case the states in a given doublet only couple to one other doublet leading to a slightly different form for the master equation as we discuss below.

To calculate the inter-doublet terms we note that close to resonance the inter-doublet transitions occur on a much larger energy scale than the spacing between levels in the doublet, $peV\gg\Delta E'$. This allows us to simplify the calculation by ignoring the $\Jp$ terms in the block-diagonal Hamiltonian [Eq.\ \eqref{eqn:defineH'}] and treat the charge states $\ket{n,k}$ as the eigenstates of the system. Thus, using the charge state basis, the inter-doublet contributions take the form
\begin{subequations}
\label{eqn:inter}
\begin{gather}
\begin{split}
 \left.\diffst{0k,0k'}\right|_{\rm inter} = \Gamma_L\tilde{\sigma}_{1k+\nicefrac{1}{2},1k'+\half} + \Gamma_R\ts{1k-\half,1k'-\half}\\-\Gamma_{{\Delta k}}\ts{0k,0k'},
\end{split} \\
 \left. \diffst{1k+\half,1k'+\half}\right|_{\rm inter}= - (\Gamma_L+\Gamma_R+\Gamma_{\Delta k})\ts{1k+\half,1k'+\half},  \\
  \left.\diffst{0k,1k'+\half}\right|_{\rm inter} = -\left(\frac{\Gamma_L+\Gamma_R}{2}+\Gamma_{\Delta k}\right) \ts{0k,1k'+\half},
\end{gather}
\end{subequations}
where we have assumed $k_{\rm B}T\ll peV$ and the dephasing rate is given by
\begin{equation}
 \Gamma_{\Delta k}=\frac{S_V(0)}{2\hbar^2}\Delta k^2,
\end{equation}
with $\Delta k=k_1-k_2$ for $\ts{ik_1,jk_2}$. The transition rates at the center of the resonance\cite{footnote2} are given by
\begin{eqnarray}
 \Gamma_L&=& \left(\frac{J}{4peV^{(p)}_{\rm res}}\right)^2\frac{S_V(\omega_p)}{\hbar^2},   \label{eqn:gamma}\\
 \Gamma_R&=& \left(\frac{J}{4(p+1)eV^{(p)}_{\rm res}}\right)^2\frac{S_V(\omega_{p+1})}{\hbar^2},   \label{eqn:gamma2}
\end{eqnarray}
with $\omega_p=2peV^{(p)}_{\rm res}/\hbar$. For the $p=0$ case there is only one decay channel linking different doublets, and the associated energy difference is $2eV$. Thus, provided $2eV\gg k_{\rm B}T,\Delta E'$ the same set of inter-doublet terms is obtained, but with $\Gamma_L=0$.

To calculate the terms in the master equation describing the intra-doublet transitions we need to use the full eigenstates of the system
[Eq. \eqref{eqn:eigenstates}] and take account of the effects of the $\Jp$ terms in the Hamiltonian. The energy differences between the states within a doublet can be much smaller than those between the doublets, and so in this case we include the effects of a finite temperature. This leads to the intra-doublet equations,
\begin{subequations}
\label{eqn:intra}
\begin{gather}
 \left.\diffst{ak,ak}\right|_{\rm intra}=\gamma_{ba}\ts{bk,bk} - \gamma_{ab}\ts{ak,ak}, \\
 \left. \diffst{bk,bk}\right|_{\rm intra}=\gamma_{ab}\ts{ak,ak} - \gamma_{ba}\ts{bk,bk},\\
 \left. \diffst{ak,bk}\right|_{\rm intra}=-\left(\frac{\gamma_{ab}+\gamma_{ba}}{2}\right)\ts{ak,bk},
\end{gather}
\end{subequations}
where $\gamma_{ij}$ gives the transition rate between the states $\ket{i,k}$ and $\ket{j,k}$ within the $k$-th doublet. These have the form
\begin{equation}
 \gamma_{ij}=\left(\frac{2p+1}{2}\right)^2\can{2}\san{2}\frac{S_V(\omega_{ij})}{\hbar^2},
\end{equation}
where $\omega_{ab}=-\omega_{ba}=-2\Delta E'/\hbar$.

Finally, we transform the inter-doublet contributions [Eq.\ \eqref{eqn:inter}] into the eigenstate basis and combine them with the intra-doublet 
terms [Eq.\ \eqref{eqn:intra}] to obtain the full master equations of the system. For sufficiently weak dissipation, $\Gamma_{L(R)}\ll |\omega_{ab}|$, it is possible to simplify the master equation significantly by applying a further rotating wave approximation (RWA). After this approximation the master equation takes the form,

\begin{subequations}
\label{eqn:ME}
\begin{multline}
  \diffst{ak,ak'}=\Gamma_{aa}^p\ts{ak_1,ak'_1} + \Gamma_{ba}^p\ts{bk_1,bk'_1} \\
+ \Gamma_{aa}^{p+1}\ts{ak_2,ak'_2} + \Gamma_{ba}^{p+1}\ts{bk_2,bk'_2}  +\gamma^{\Delta k}_{ba}\ts{bk,bk'} \\ - (\Gamma_{ab}^p+\Gamma_{aa}^{p}+\Gamma_{ab}^{p+1}+\Gamma_{aa}^{p+1}+\gamma^{\Delta k}_{ab}+\Gamma_{\Delta k})\ts{ak,ak'}, \label{eqn:MEaa}
\end{multline}
\begin{multline}
  \diffst{bk,bk'}=\Gamma_{ab}^{p}\ts{ak_1,ak'_1} + \Gamma_{bb}^p\ts{bk_1,bk'_1}  \\
+ \Gamma_{ab}^{p+1}\ts{ak_2,ak'_2} + \Gamma_{bb}^{p+1}\ts{bk_2,bk'_2}  +\gamma^{\Delta k}_{ab}\ts{ak,ak'} \\ -(\Gamma_{ba}^{p}+\Gamma_{bb}^{p}+\Gamma_{ba}^{p+1}+\Gamma_{bb}^{p+1}+\gamma^{\Delta k}_{ba}+\Gamma_{\Delta k}) \ts{bk,bk'}, \label{eqn:MEbb}
\end{multline}
\begin{multline}
 \diffst{ak,bk'}=-\left(\frac{\Gamma_{aa}^p+\Gamma_{bb}^p}{2}\right)\ts{ak_1,bk'_1} \\
 -\left(\frac{\Gamma_{aa}^{p+1}+\Gamma_{bb}^{p+1}}{2}\right)\ts{ak_2,bk'_2} \\ -\left(\frac{\Gamma_L+\Gamma_R}{2}+\gamma^{\Delta k}_{\rm coh}+\Gamma_{\Delta k}\right)\ts{ak,bk'}, \label{eqn:MEab}
\end{multline}
\end{subequations}
where $k_1=k-p$, $k_2=k-p-1$ and $\Gamma_{ij}^p$ are the transition rates between the states $\ket{i,k+p}\rightarrow\ket{j,k}$, which are given by
\begin{subequations}
\begin{gather}
 \Gamma_{aa}^p=\Gamma_{bb}^p=\Gamma_L\cos^2\alpha\sin^2\alpha, \\   \Gamma_{aa}^{p+1}=\Gamma_{bb}^{p+1}=\Gamma_R\cos^2\alpha\sin^2\alpha, \\
 \Gamma_{ba}^{p}=\Gamma_L\cos^4\alpha, \\  \Gamma_{ba}^{p+1}=\Gamma_R\cos^4\alpha,  \\
 \Gamma_{ab}^p=\Gamma_L\sin^4\alpha, \\   \Gamma_{ab}^{p+1}=\Gamma_R\sin^4\alpha.
\end{gather}
\end{subequations}
We have also defined
\begin{widetext}
\begin{gather}
 \gamma_{ij}^{\Delta k}=\left\{ \begin{array}{lr} \gamma_{ij}, & \Delta k=0, \\ \left(\frac{2p+1}{2}\right)^2\san{2}\can{2}\frac{S_V(0)}{\hbar^2}, & \Delta k\neq 0, \end{array} \right. \\
\gamma^{\Delta k}_{\rm coh}=\left\{ \begin{array}{lr} \frac{\gamma_{ab}+\gamma_{ba}}{2}, & \Delta k=0, \\ \left(\frac{2p+1}{2}\right)\left[\left(\frac{2p+1}{2}\right)(\can{4}+\san{4})+\Delta k\cos 2\alpha\right] \frac{S_V(0)}{\hbar^2}, & \Delta k\neq 0.\end{array}\right.
\end{gather}
\end{widetext}

For the RWA to be valid we require that incoherent decay rates are much smaller than the Bohr frequency associated with the doublets, 
which, for $p\geq1$, results in the condition $\Gamma_L\ll 2|J_p|/\hbar$ (since $\Gamma_L$ is the largest decay rate). 
From Eqs.\ \eqref{eqn:jp} and \eqref{eqn:gamma}, we find that, $\hbar\Gamma_L/|J_p|\propto (R/R_Q)(eV^{(p)}_{\rm res}/J)^{2p-1}$. By tuning the gate 
voltage, $V^{(p)}_{\rm res}$ can take on any value in the range $0<(2p+1)eV^{(p)}_{\rm res}<E_C$, thus, for a given value of $R$, the requirement that the RWA is valid puts a limit (which becomes stricter as $p$ increases) on the maximum voltage that can be considered. Our interest here is in the regime where the SSET charge dynamics is largely coherent, leading to sharp resonances in the current, and hence we naturally focus on the regime where the RWA is valid. Since these conditions can only be met in practice\cite{leppakangas:08,haviland:94,billangeon:07} for the lower values of $p$, we will concentrate on the $p=0,1$ resonances.

The master equations bear a strong resemblance to those which describe the radiative cascade of quantum optics.\cite{cohen:92} In the radiative cascade a laser field drives a two-level atom and when the field is treated quantum mechanically the eigenstates of the system are atom-field hybrids (dressed states). Decay processes lead to a cascade in which photons are emitted and the laser state moves progressively towards lower photon numbers. Analogously in the SSET the states $\left\{|a,k\rangle,|b,k\rangle\right\}$ are like the atom-laser dressed states, with the island charge states playing the role of the atom and $k$  like the state of the laser. In this case decay processes generate a cascade in which the system evolves towards states of ever increasing $k$.

\subsection{Effective two-level system}

Although the full master equation for the SSET is rather complicated, it is possible to derive a much simpler set by tracing over the charges that have passed through the SSET.
Defining a set of reduced coherences,\cite{cohen:92}
\begin{equation}
\rho^q_{ij}(t)=\sum_k\bra{i,k+q}\sigma(t)\ket{j,k},
\end{equation}
and carrying out the trace over $k$, we obtain a much simpler matrix equation,
\begin{equation}
 \dot{\vector{\rho}}^q(t)=(2iqeV+\mat{M})\vector{\rho}^q(t), \label{eqn:TLS}
\end{equation}
where $\vector{\rho}^q=(\rho_{aa}^q, \rho_{bb}^q, \rho_{ab}^q)^T$.  Within the RWA the form of the matrix $\mat{M}$ is given by
\begin{equation}
 \mat{M}= \begin{pmatrix}
     -(\Gamma_q+\Gamma^q_a) & \Gamma^q_b & 0 \\
     \Gamma^q_a & -(\Gamma_q+\Gamma^q_b) & 0 \\
     0 & 0 & -(i\omega_{ab}+\Gamma^q_{\text{coh}})
    \end{pmatrix},
\end{equation}
where $\Gamma_q=\Gamma_{\Delta k=q}$ and the other rates are given by,
\begin{gather}
  \Gamma_a^q=(\Gamma_L+\Gamma_R)\can{4}+\gamma^q_{ba}, \\
  \Gamma_b^q=(\Gamma_L+\Gamma_R)\san{4}+\gamma^q_{ab},\\
  \Gamma_\text{coh}^q=\frac{(\Gamma_L+\Gamma_R)(1+2\can{2}\san{2})}{2}+\gamma_{\rm coh}^q+\Gamma_{q}. \label{eqn:gcoh}
\end{gather}
By taking $q=0$ we obtain the equation which describes the evolution of the island charge. The resulting master equation describes a two-level system (TLS) which is both driven and damped. This simple two-level description is all that is required to calculate the main charge noise properties of the SSET.

 The full density operator does not have a steady state, the system cascades to increasing values of $k$ as charge tunnels though the transistor. The reduced equations derived above, however, do have a well defined steady state.  All of the reduced coherences where $q\neq 0$ are zero in the steady state, $\steady[q\neq0]{ij}=0$. The $q=0$ case has the steady state:  $\steady{aa(bb)}=\Gamma_{b(a)}/(\Gamma_a+\Gamma_b)$, and $\steady{ab}=0$. Note that here and in what follows  we drop the superscripts on $\rho$ and $\Gamma$ for the case  $q=0$.


\section{Average current}
\label{sec:current}

\begin{figure}
 \center{  \epsfig{file=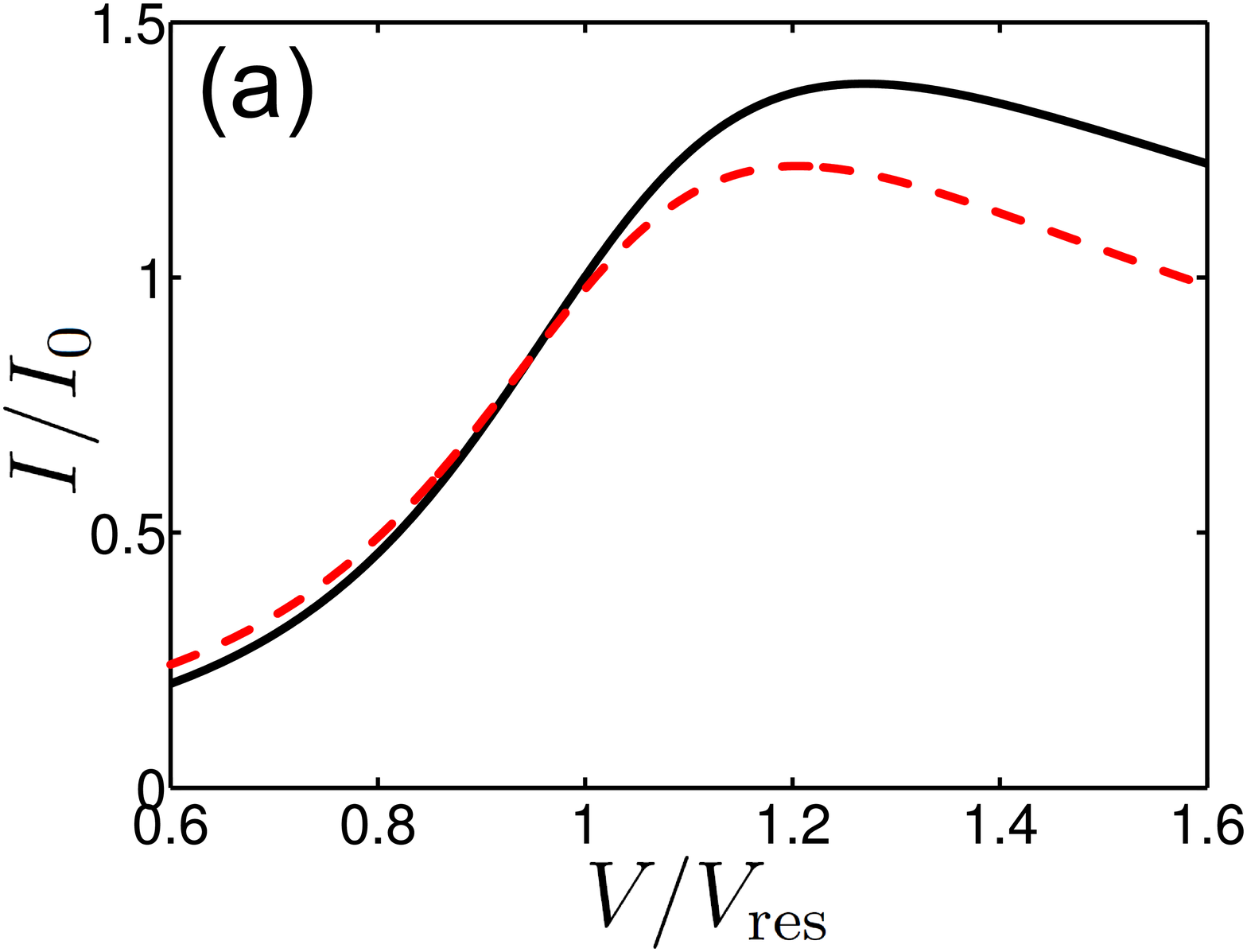, width=0.35\textwidth}
           \epsfig{file=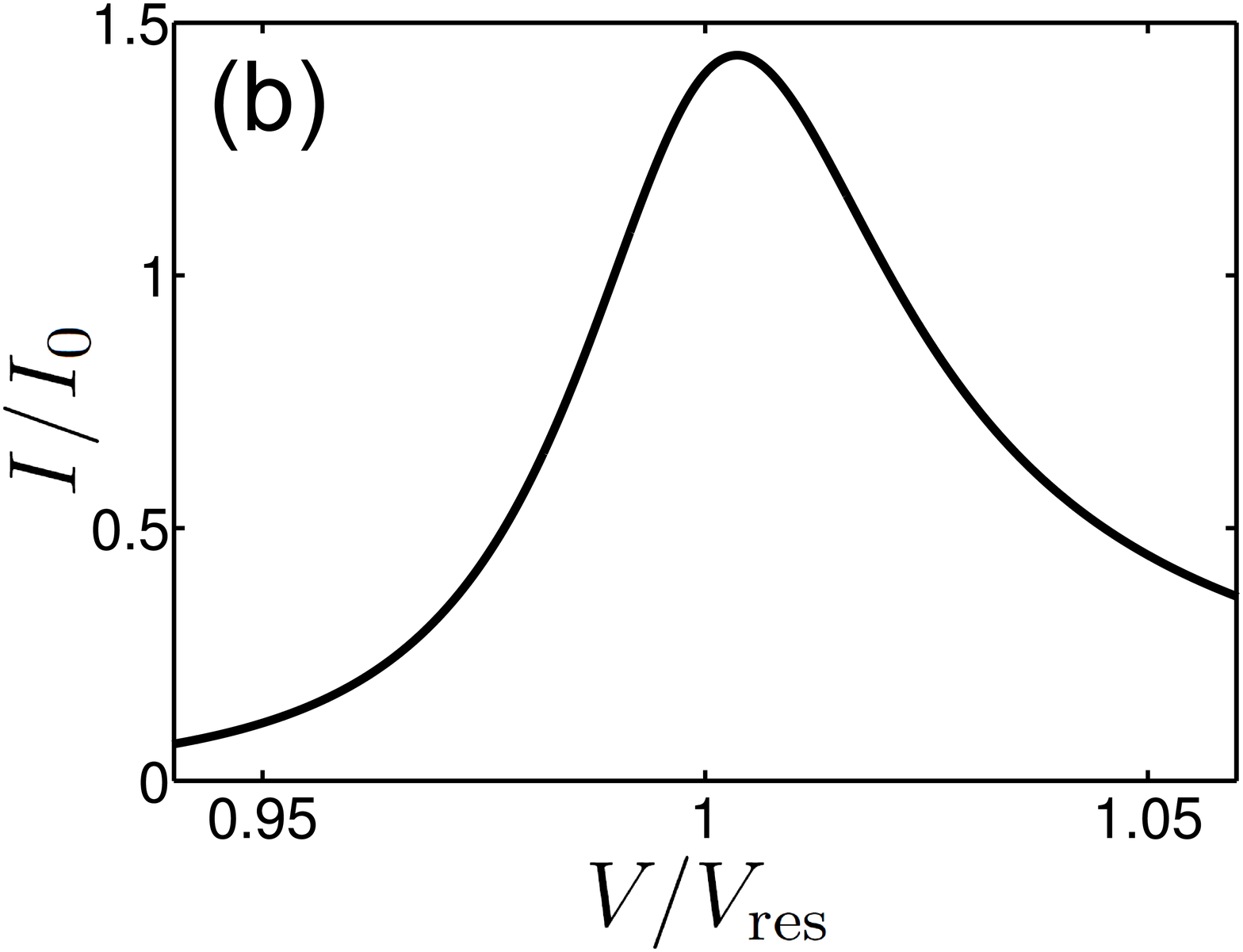 , width=0.35\textwidth}}
            \caption{(Color online) Current-voltage characteristics of the SSET in the vicinity of (a) the $p=0$ resonance (b) the $p=1$ resonance.
            The current is scaled by $I_0=e(\Gamma_L+\Gamma_R)$ in each case. The values of the other parameters are given in the text.
            In (a) we also show in the dashed, red line the current calculated using the full voltage dependence of the
            decay rate $\Gamma_R$ and the Hamiltonian.}\label{fig:current}
\end{figure}

As a first application of the master equation \eqref{eqn:ME}, we calculate the steady state average current, $\Iss$, through the transistor.  The current is determined by the rate of change of the number of Cooper-pairs which have crossed the device,
\begin{equation}
  \Iss=2e\diff{\average {k'}}{t}=2e\Tr[k'\dot\sigma].
\end{equation}
We calculate only the dominant term which comes from the lowest order part of the $k'$ expansion, $\kn{0}=k$, as the next 
lowest order contribution (from $k^{(1)}$) vanishes 
and we neglect higher order contributions.
 The only non-zero terms after performing the trace come from the dissipative parts of the diagonal master equations 
[Eqs.\ \eqref{eqn:MEaa}--\eqref{eqn:MEbb}] which give rise to the incoherent transitions. Since these terms only depend on the
 reduced coherences of the system, the current reaches a well-defined steady-state in the long time limit,
\begin{multline}
  \Iss=2e\left(\steady{aa}\san{2}+\steady{bb}\can{2}\right)\\ \times \left(p\Gamma_L+(p+1)\Gamma_R\right). \label{eqn:Ibar}
\end{multline}
Within the regime where the RWA is valid, this matches the results obtained previously\cite{maassen:91a,joyez:95, siewert:96} using a rate equation approach and Fermi's golden rule.

The current near the $p=0$ and $p=1$ resonances is shown in Fig.\ \ref{fig:current} for the typical parameter
values\cite{leppakangas:08,haviland:94,billangeon:07} $E_C=4E_J=100$\,$\mu$eV, $T=30$\,mK, a resistance for the embedding
circuit $R=50$\,$\Omega$ and we have set $n_g=0.1$ [we use these parameters throughout for the numerical calculations].
The full lines in Fig.\ \ref{fig:current} show the current calculated with the decay rates, $\Jp$ and the second order correction to $\Delta E$ given by their on-resonance values.\cite{footnote3,footnote2}
As the  $p=0$ resonance is very broad (in comparison to the $p=1$ resonance), we have also
calculated the current including the full voltage dependence of the relevant decay rate and the Hamiltonian [dashed curve in Fig.\ \ref{fig:current}(a)].

The two resonances show rather different characteristics. The current around the $p=0$ resonance is broad and highly asymmetric. This is because 
in this case purely dissipative processes can generate a dc current for $V>V_{\rm res}$ (incoherent Cooper-pair tunneling\cite{nazarov:09}) as can be 
seen from the energy level diagram in Fig.\ \ref{fig:presonances}(b). The current for the $p=1$ resonance is much closer to the standard 
Lorentzian form of a resonance, the small amount of asymmetry still present arises from the intra-doublet transitions. For $V<V_{\rm res}$, relaxation 
between the levels of the doublets (controlled by $\gamma_{ab(ba)}$) hinders current flow whilst for $V>V_{\rm res}$ it helps it. This leads to a small 
asymmetry in the current as a function of voltage which is only removed when the temperature is sufficiently high such that $\gamma_{ab}\simeq \gamma_{ba}$ 
and a Lorentzian shape is recovered.

Extending our calculation to include the regime where the RWA is no longer valid, we find that  the current peaks at the resonances become suppressed. This is because outside of the RWA the system is unable to build up the coherence between charge states necessary for current to flow. This is consistent with Ref.\ \onlinecite{leppakangas:08} where this effect was studied in detail.

\section{Charge Noise Spectrum}
\label{sec:CN}

We now turn to consider the noise properties of the SSET close to the Cooper-pair resonances. The charge noise spectrum provides a fingerprint of the subtle interplay of coherent dynamics and dissipation in the system whilst also controlling the back-action that the SSET will exert when it is used as a measuring device.

The charge noise spectrum is given by,
\begin{equation}
  S_{nn}(\omega)=\int_{-\infty}^{\infty}\average{\delta n'(t)\delta n'(0)}{\rm e}^{i\omega t}dt,
\end{equation}
where $\delta n'=n'-\average{n'}$ is the fluctuating part of the transformed island charge operator and the averages are taken over the steady state. The operator $n'$ can be written as a series expansion in terms of powers of $J$ (as we did for $k'$), $n'=n^{(0)}+n^{(1)}+\ldots$, as described in the appendix. The dominant contribution to the spectrum comes from the zeroth order terms in the expansion, $n^{(0)}= n$, and so we consider these terms first. The higher order terms in $n'$ give rise to weaker features in the spectrum which we go on to calculate in Sec.\ \ref{subsec:HighOrder}.

\subsection{Triplet structure}
\begin{figure}
 \center{\epsfig{file=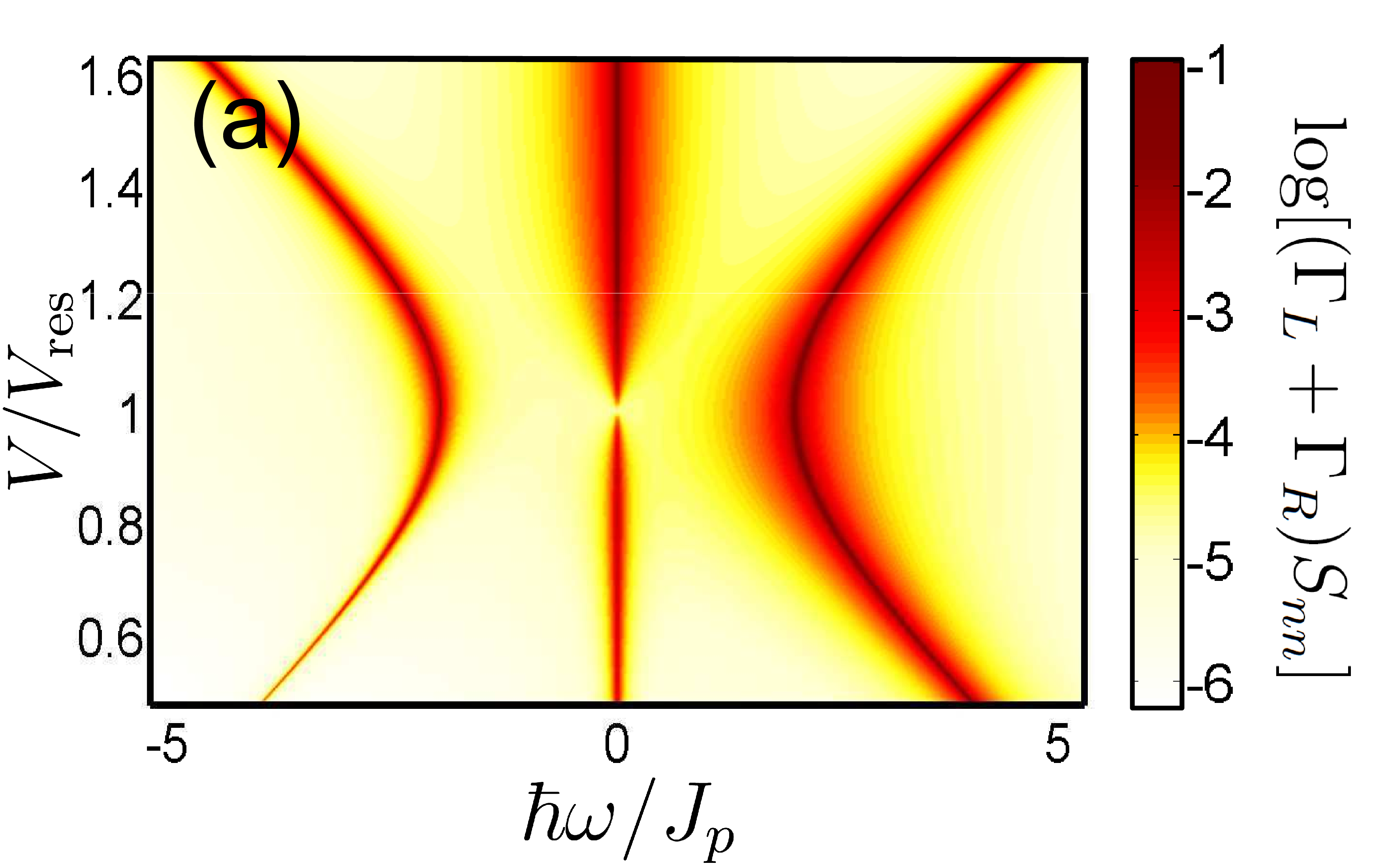, width=0.4\textwidth}
         \epsfig{file=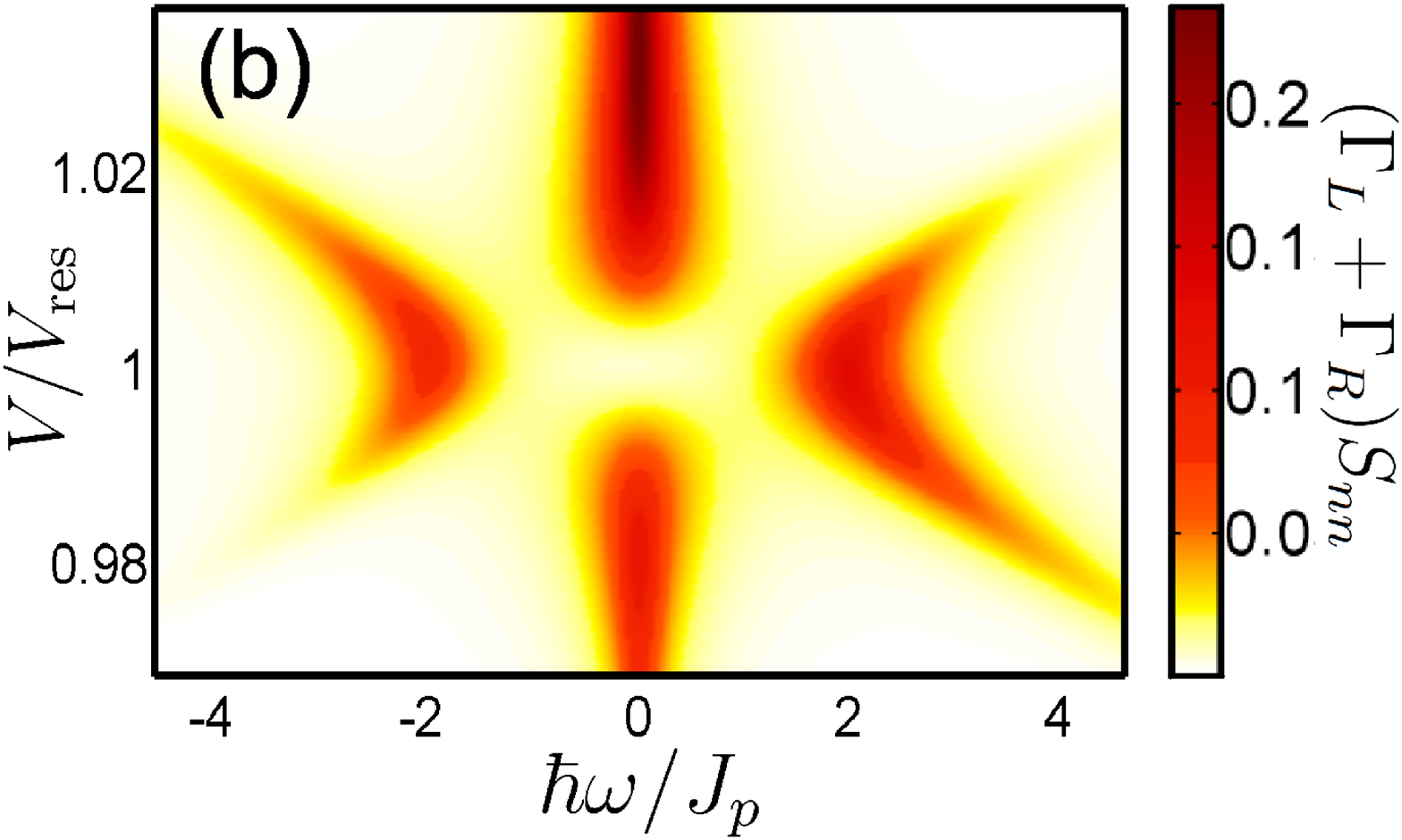, width=0.4\textwidth}}
 \caption{(Color online) Charge noise spectrum as a function of frequency and voltage around the (a) $p=0$ and (b) $p=1$ resonances. The resonances occur at the point where the central peak disappears.}\label{fig:CNomega}
\end{figure}

The charge noise spectrum is readily obtained using the quantum regression theorem\cite{cohen:92} which allows us to calculate the behavior of $\average{n(t)n(0)}$ for $t>0$ by using the equation of motion for $\average{n(t)}$ with appropriately modified initial conditions. The starting point for the calculation is the effective master equation for the SSET charge state [Eq.\ \eqref{eqn:TLS}] with $q=0$.

We begin by defining a series of projection operators
\begin{equation}
  p_{ij}=\sum_k \ket{j,k}\bra{i,k},
\end{equation}
with the property $\average{p_{ij}}=\steady{ij}$  which then allow us to calculate the fluctuating part $\delta p_{ij}=p_{ij}-\steady{ij}$. The
 correlation function for the charge written in terms of the projectors is,
\begin{equation}
  \average{\delta n(t)\delta n(0)}=\sum_{i,j,l,m} n_{ij}n_{lm}\average{\delta p_{ij}(t)\delta p_{lm}(0)},
\end{equation}
where the matrix elements are given by the transformation of $n$ into the eigenstate basis, $n_{aa}=\san{2}$, $n_{bb}=\can{2}$ and $n_{ab}=n_{ba}=\sa\ca$.

Using the regression theorem and the initial conditions for the projectors,
\begin{equation}
  \average{\delta p_{ij}(0)\delta p_{lm}(0)}=\delta_{im}\steady{jl}-\steady{ij}\steady{lm},
\end{equation}
we obtain the spectrum,
\begin{multline}
S_{nn}(\omega)=2\steady{aa}\steady{bb}(n_{aa}-n_{bb})^2\frac{\Gamma_{\text{pop}}}{\omega^2+\Gamma_{\text{pop}}^2} \\+2n_{ab}^2\left(\frac{\steady{bb}\Gamma_{\text{coh}}}{(\omega-\omega_{ab})^2+\Gamma_{\text{coh}}^2}  +\frac{\steady{aa}\Gamma_{\text{coh}}}{(\omega-\omega_{ba})^2+\Gamma_{\text{coh}}^2}\right), \label{eqn:chargenoiseeigen}
\end{multline}
where $\Gamma_{\rm pop}=\Gamma_a+\Gamma_b$. Examples of the spectrum as a function of frequency and voltage around the $p=0$ and $p=1$ resonances are shown in Fig.\ \ref{fig:CNomega}. In both cases the spectrum consists of three Lorentzians. This triplet structure is exactly what is expected for a coherent TLS in the presence of  dissipation.\cite{cohen:92,jaehne:08} The central peak around zero frequency arises due to incoherent transitions between eigenstates, its height is determined by $|n_{aa}-n_{bb}|$, the difference in average charge between the $\ket{a}$ and $\ket{b}$ eigenstates. This peak disappears when the system is tuned to resonance since the eigenstates are equal mixtures of charge states.

The sidepeaks arise at $\omega=\pm \omega_{ab}$ due to coherent oscillations between the eigenstates. The heights of the
sidepeaks are controlled by the steady state populations of the eigenstates, which is what provides the asymmetry between the
negative and positive frequency peaks, and the width is given by the rate at which the coherences decay, $\Gamma_{\text{coh}}$
[see Eq.\ \eqref{eqn:gcoh}]. Note that the spacing of the sidepeaks is much larger for $p=0$, the splitting $J_{p=0}\sim 25$\,GHz, compared to
$J_{p=1}\sim 1$\,GHz for $p=1$. This means that finite temperature effects are not as important in the $p=0$ resonance.

\subsection{Back-action}

\begin{figure}
 \center{\epsfig{file=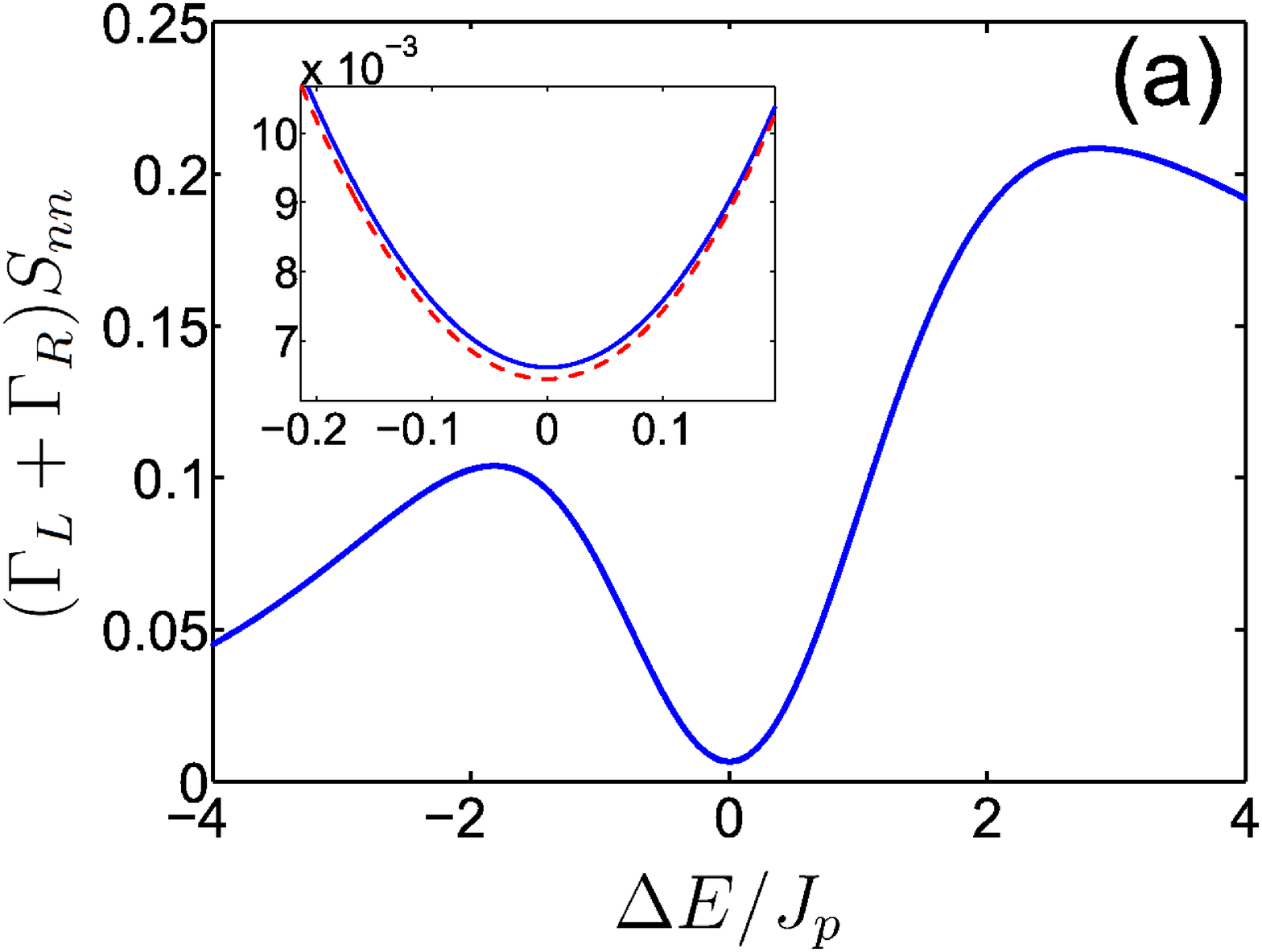, width=0.35\textwidth}
         \epsfig{file=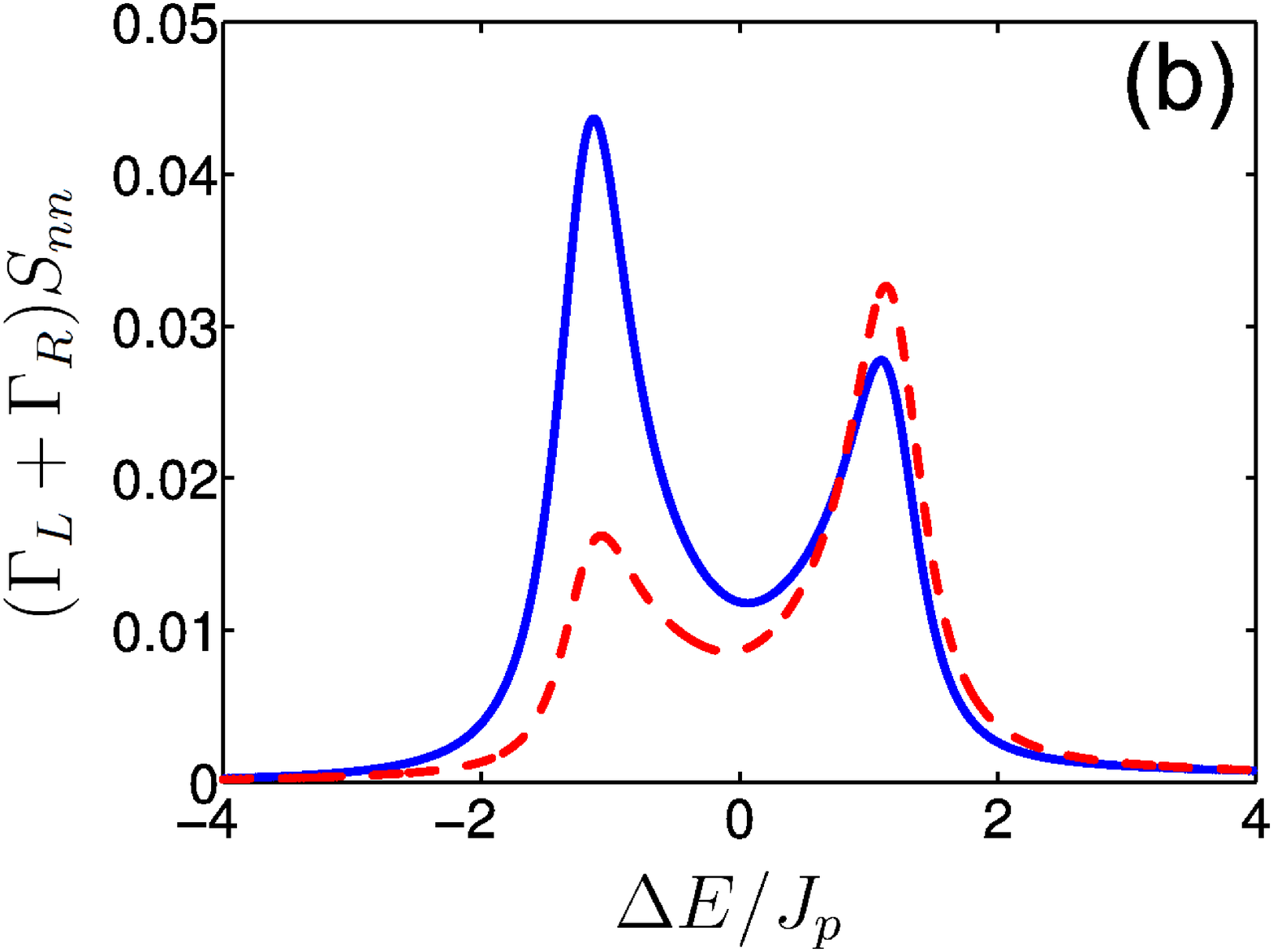, width=0.35\textwidth}}
 \caption{(Color online) Charge noise, $S_{nn}$, as a function of $\Delta E$ for (a) $\Omega=100$\,MHz and (b) $\Omega=4$\,GHz oscillator. Solid (blue) lines are $S_{nn}(\Omega)$, dashed (red) lines are $S_{nn}(-\Omega)$. The inset in (a) shows the region around the minimum where the difference between the two curves is most pronounced.}\label{fig:Snn}
\end{figure}

We now consider the effect that the charge noise spectrum has on another system coupled to the SSET island. For concreteness
we consider the case of a resonator with a charge-position coupling,
\begin{equation}
  H_{\text{c}}=\lambda n(c+c^{\dagger}),
\end{equation}
where $c$ is the resonator lowering operator, and $\lambda$ the coupling strength. For weak coupling, and provided that the resonator does not start to self-oscillate,\cite{clerk:05,rodrigues:07} linear response theory can be used to calculate the effect of the SSET on the resonator dynamics.\cite{clerk:04,clerk:08} This back-action of the SSET on a resonator can be useful, providing in some cases a way to cool the resonator\cite{clerk:05,blencowe:05} whilst also providing a way of measuring the asymmetry in the SSET's noise properties.\cite{xue:09}

Within the linear response regime, the SSET acts on the resonator like an additional thermal bath and its effect can be characterized by an effective damping, $\gba$, and an effective thermal occupation number, $\nba$. For a weakly damped resonator of frequency $\Omega$, these are given by\cite{marquardt:07,clerk:08}
\begin{gather}
  \gba(\Omega)=\lambda^2\left[S_{nn}(\Omega)-S_{nn}(-\Omega)\right], \\
 \nba(\Omega)=\left[\frac{S_{nn}(\Omega)}{S_{nn}(-\Omega)}-1\right]^{-1}.
\end{gather}
Measurement of $\gba$ and  $\nba$ allows both frequency components of the charge noise spectrum, $S_{nn}(\pm\Omega)$, to be inferred. We will explore
the back-action effect of the SSET on resonators in two different regimes of frequency, $\Omega=100$\,MHz and $\Omega=4$\,GHz, corresponding to
different physical realizations of the resonator. The $100$\,MHz frequency is typical of a nanomechanical resonator\cite{knobel:03} whilst
the $4$\,GHz frequency matches that of superconducting striplines\cite{wallraff:04} or LC resonators.\cite{xue:09} We focus on the $p=1$ resonance  from now on since the sidepeaks of the $p=0$ resonance are too wide to have any significant effect on even a high frequency LC oscillator.

Figure \ref{fig:Snn} shows the charge noise as a function of the detuning from resonance for $\Omega=\pm100$\,MHz and $\Omega=\pm 4$\,GHz. At $100$\,MHz the spectrum is very symmetric, $S_{nn}(\Omega)\approx S_{nn}(-\Omega)$ because the sidepeaks have very little weight at this frequency. In this case maximum asymmetry is achieved at the center of the resonance ($\Delta E=0$) when the spacing of the outer peaks is minimized. In contrast, at $4$\,GHz the spectrum is highly asymmetric as here the sidepeaks cross through this frequency.

The curves for $S_{nn}(\pm\Omega)$ are not simply reflections of each other, as would be expected for a classically driven
TLS\cite{jaehne:08} or other resonances in the SSET.\cite{clerk:05,blencowe:05} This asymmetry occurs for the same reason that the current peak
is not a simple Lorentzian; the intra-doublet decay rates are not symmetric between $\pm\Delta E$. Similar effects are seen in a driven TLS when\cite{wang:09} the temperature dependence of the  relaxation rate is taken into account. The behavior of the driven TLS where $S_{nn}(\Omega, \Delta E)=S_{nn}(-\Omega, -\Delta E)$ is recovered in the high temperature limit where the intra-doublet transition is saturated, $\gamma_{ab}=\gamma_{ba}$. We can understand how the intra-doublet decays lead to this asymmetric behavior in terms of effective temperatures. The intra-doublet decays drive the eigenstate populations towards an equilibrium distribution which corresponds to the temperature of the bath, this is constant and always positive. However, the inter-doublet rates drive the SSET to an equilibrium point whose effective temperature varies strongly with $\Delta E$ and, in particular, changes sign when $\Delta E=0$. The competition between these two behaviors causes the asymmetry in this system.\cite{clerk:05,blencowe:05}

\begin{figure*}
 \center{\epsfig{file=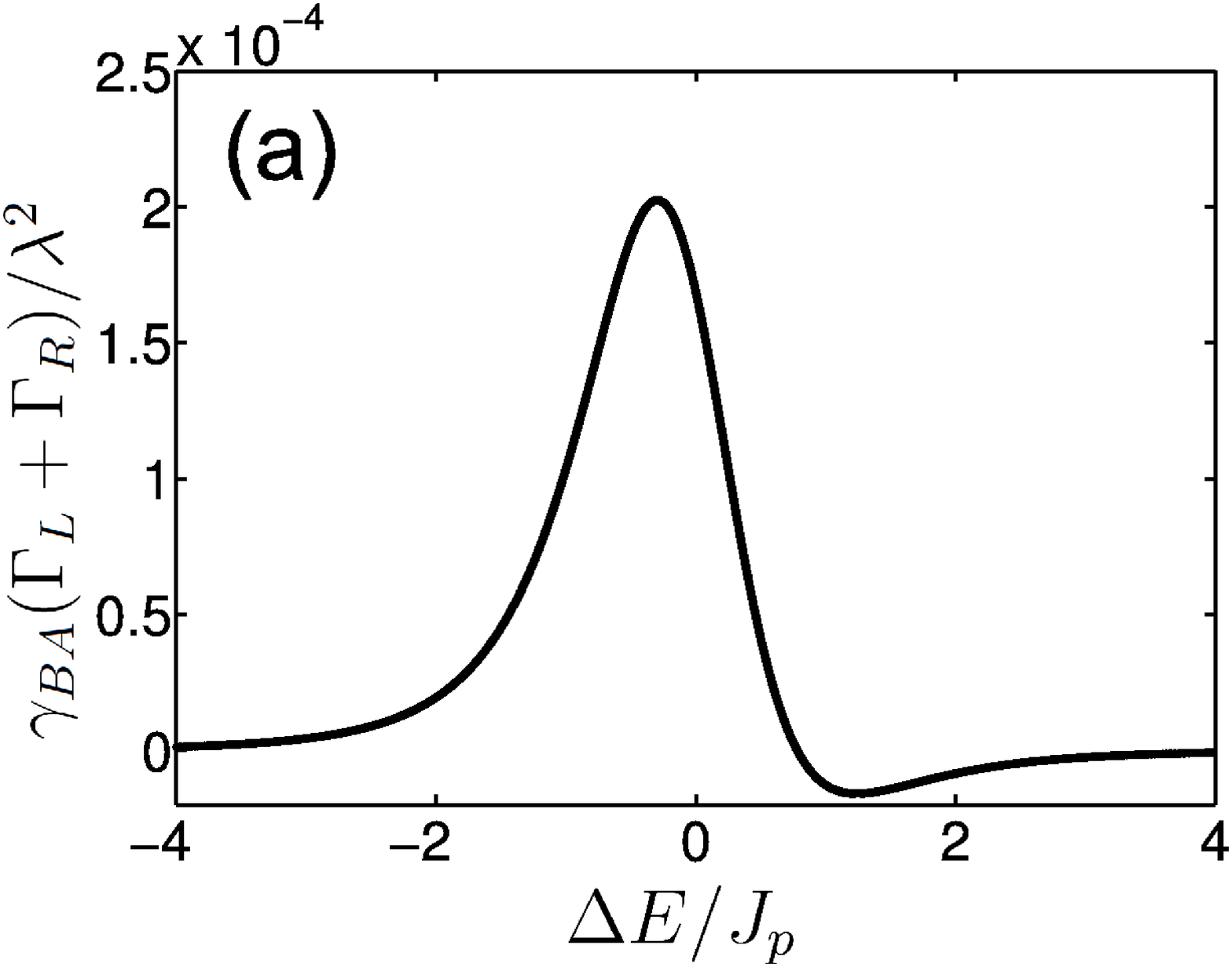, width=0.35\textwidth}
        \epsfig{file=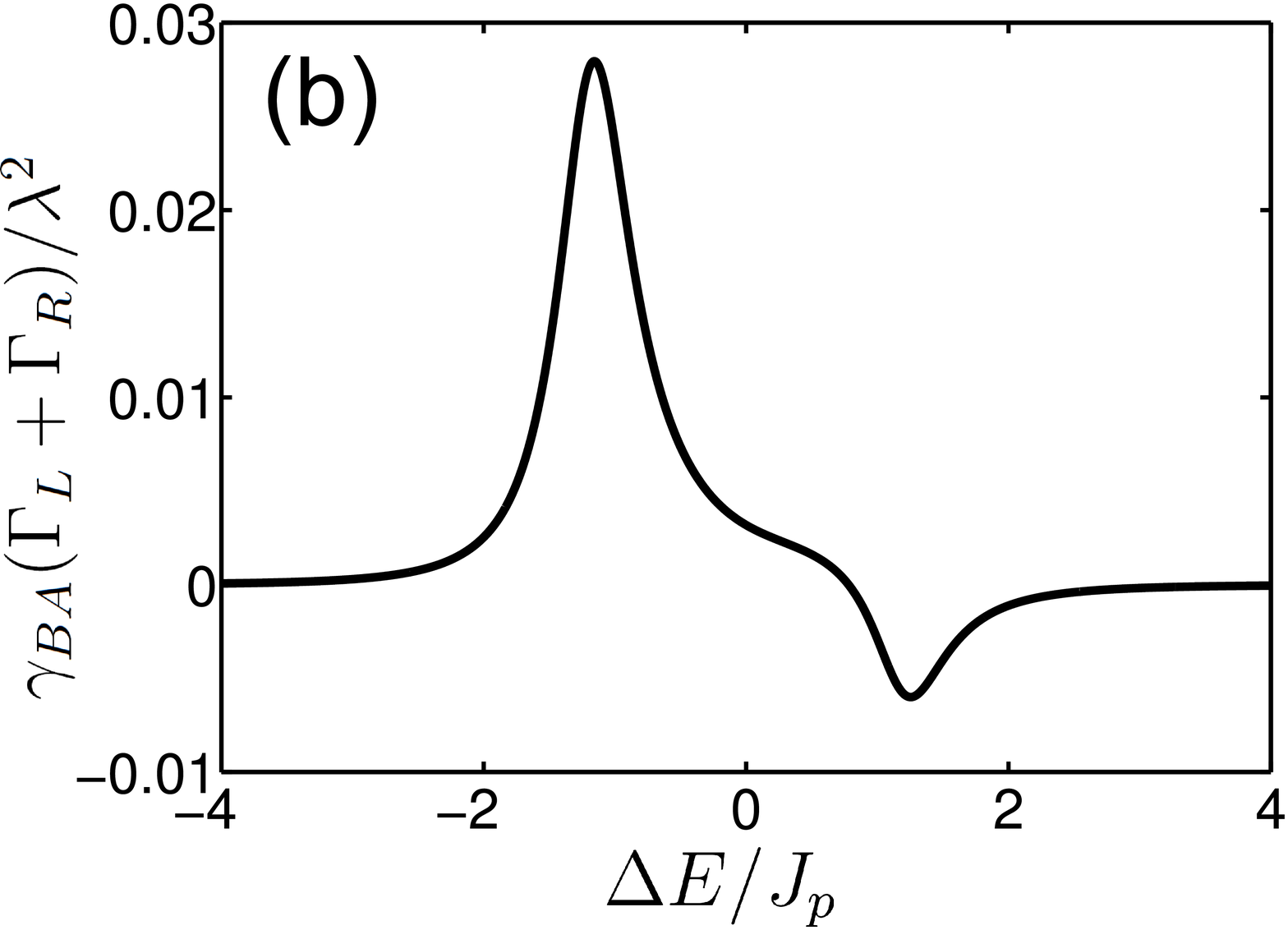, width=0.35\textwidth}
         \epsfig{file=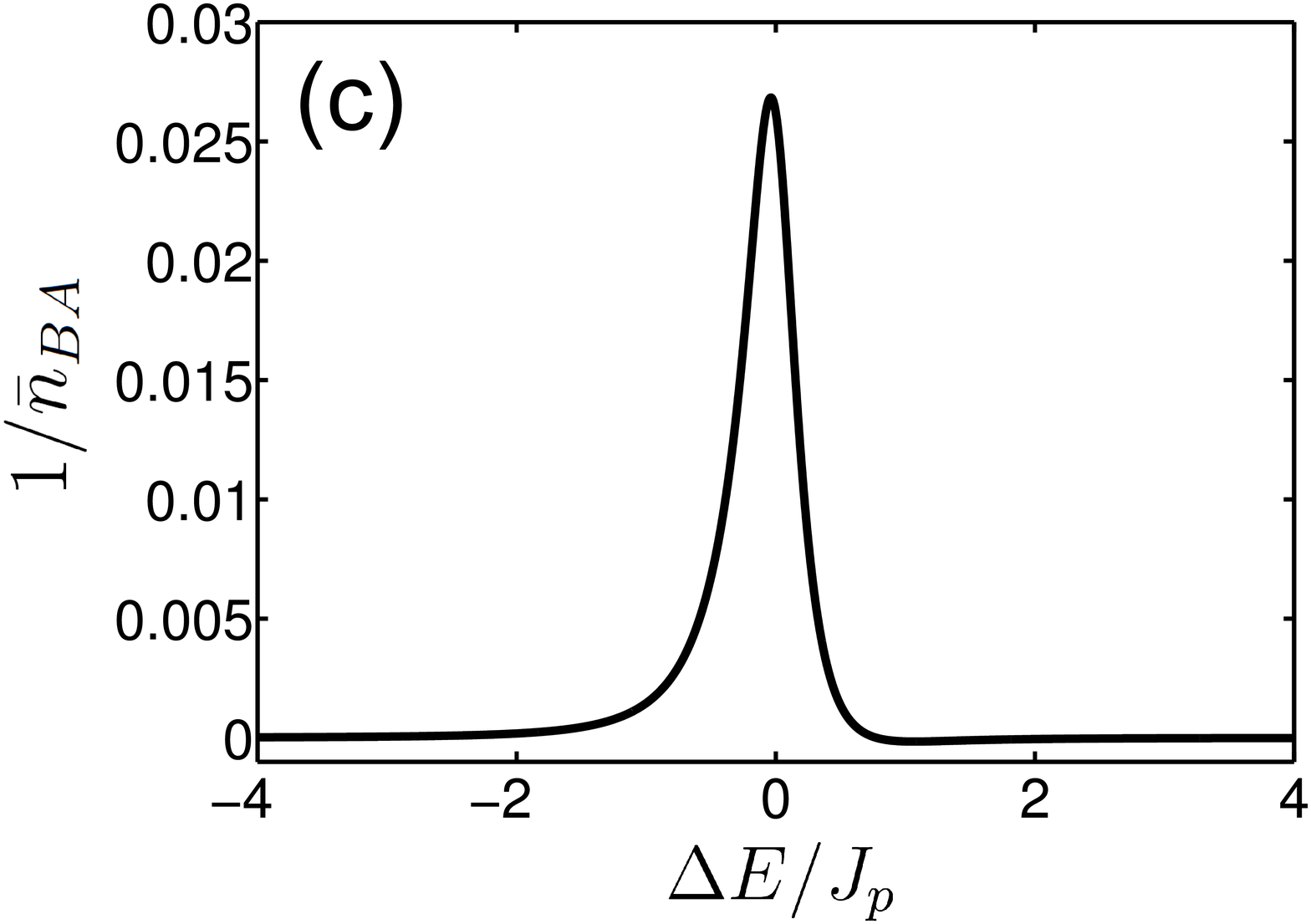, width=0.35\textwidth}
         \epsfig{file=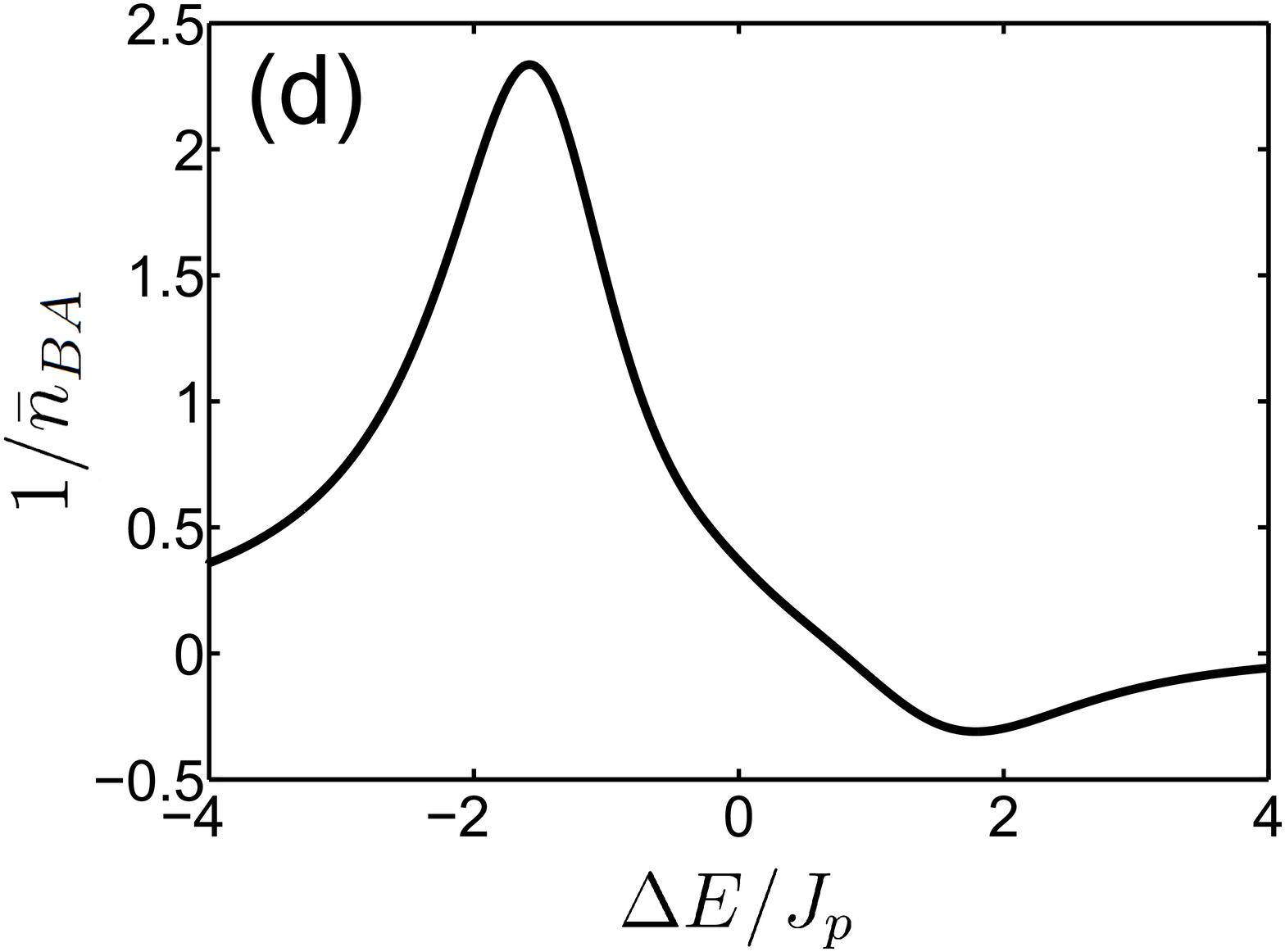, width=0.35\textwidth}}
 \caption{Back-action damping,  $\gba$ [(a)-(b)] and  effective occupation number, $\nba$ [(c)-(d)] of an oscillator weakly coupled to the SSET. (a) and (c) are for  $\Omega=100$\,MHz, (b) and (d) for $\Omega=4$\,GHz.}\label{fig:gammanbar}
\end{figure*}

In Fig.\ \ref{fig:gammanbar} we plot the damping, $\gba$, for $\Omega=100$\,MHz and $\Omega=4$\,GHz. The small asymmetry in the low frequency
noise spectrum gives the damping a very small magnitude, but for the high frequency resonator it is much larger. The lack of symmetry in
the noise spectrum, $S_{nn}(\Omega, \Delta E)\neq S_{nn}(-\Omega, -\Delta E)$, leads to quite different magnitudes for the damping at $\pm\Delta E$, with the  anti-damping peak suppressed by the intra-doublet decays for both the high and low frequency cases.

When a resonator is coupled to the SSET its steady state is determined by a combination of the back-action of the SSET and the influence 
of the rest of the resonator's surroundings which are in thermal equilibrium at a temperature $T$, and give rise to a damping rate 
$\gamma_{\rm ext}$. The overall occupation number of the resonator, ${\overline{n}}_{\rm tot}$, is given by the average,\cite{marquardt:07,wilson:07,jaehne:08}
 \begin{equation}
 {\overline{n}}_{\rm tot}=\frac{\gamma_{\rm ext}{\overline{n}}+\gamma_{\rm BA}\nba}{\gamma_{\rm ext}+\gamma_{\rm BA}},
\end{equation}
where $\overline{n}=(\exp(\hbar\Omega/k_BT)-1)^{-1}$. Thus at a given $T$, the SSET can be used to cool the resonator provided
provided $\nba< \overline{n}$. Such cooling is important for nanomechanical resonators with frequencies in the MHz range as even at temperatures $T$ below $100$\,mK they will still contain a large number of thermal quanta.\cite{naik:06,marquardt:07,wilson:07,jaehne:08}

Figures \ref{fig:gammanbar}(c) and \ref{fig:gammanbar}(d) show the behavior of $\nba$ (plotted as $1/\nba$ to emphasize the behavior near minimas) around the resonance at $100$\,MHz and $4$\,GHz. For $\Omega=100$\,MHz, the minimum in $\nba$ occurs when the system is tuned directly to resonance where the central peak in the noise spectrum vanishes. For the typical device parameters we have chosen (given in Sec.\ \ref{sec:current}), we find a minimum of $\nba\approx 37$ which corresponds through the expression $\nba=(\exp(\hbar\Omega/k_BT_{BA})-1)^{-1}$ to an effective temperature  $T_{BA}\approx 29$\,mK, only slightly lower than the bath temperature, $T=30$\,mK. This result is, however, dependent on $T$ through the intra-doublet rates and the relative cooling potential does improve at higher bath temperatures, for example $T_{BA}\approx 37$\,mK for $T=50$\,mK and $T_{BA}\approx50$\,mK for $T=80$\,mK.

The main problem with using the SSET tuned to a Cooper-pair resonance to cool a mechanical resonator is the spacing of the peaks in the noise spectrum. For effective cooling, we need the frequency of the resonator to match the separation of the peaks. This cannot be achieved at low
frequencies in this device since the minimum intra-doublet spacing (and hence minimum peak splitting) $2\Jp/\hbar$ is in the GHz range. Trying to engineer a device where this splitting was much smaller would lead to a deterioration in the effectiveness of the cooling because of the effect of thermal noise on the SSET itself: the asymmetry in the peaks would be reduced and they would also be broadened (this would take the system outside both  the regime where the RWA is valid and also the resolved sideband limit where optimal cooling can be achieved\cite{marquardt:07,wilson:07}). In contrast, for a driven TLS\cite{jaehne:08} the potential for cooling is much greater as it is possible to tune the drive (which corresponds to the Josephson coupling in our system) and it is the {\it difference} between this frequency and the level separation of the TLS(both of which can be $\gg k_{\rm B}T/\hbar$) that sets the spacing of the sidepeaks in the corresponding noise spectrum.

\subsection{Higher order spectral features}
\label{subsec:HighOrder}

\begin{figure*}
          \epsfig{file=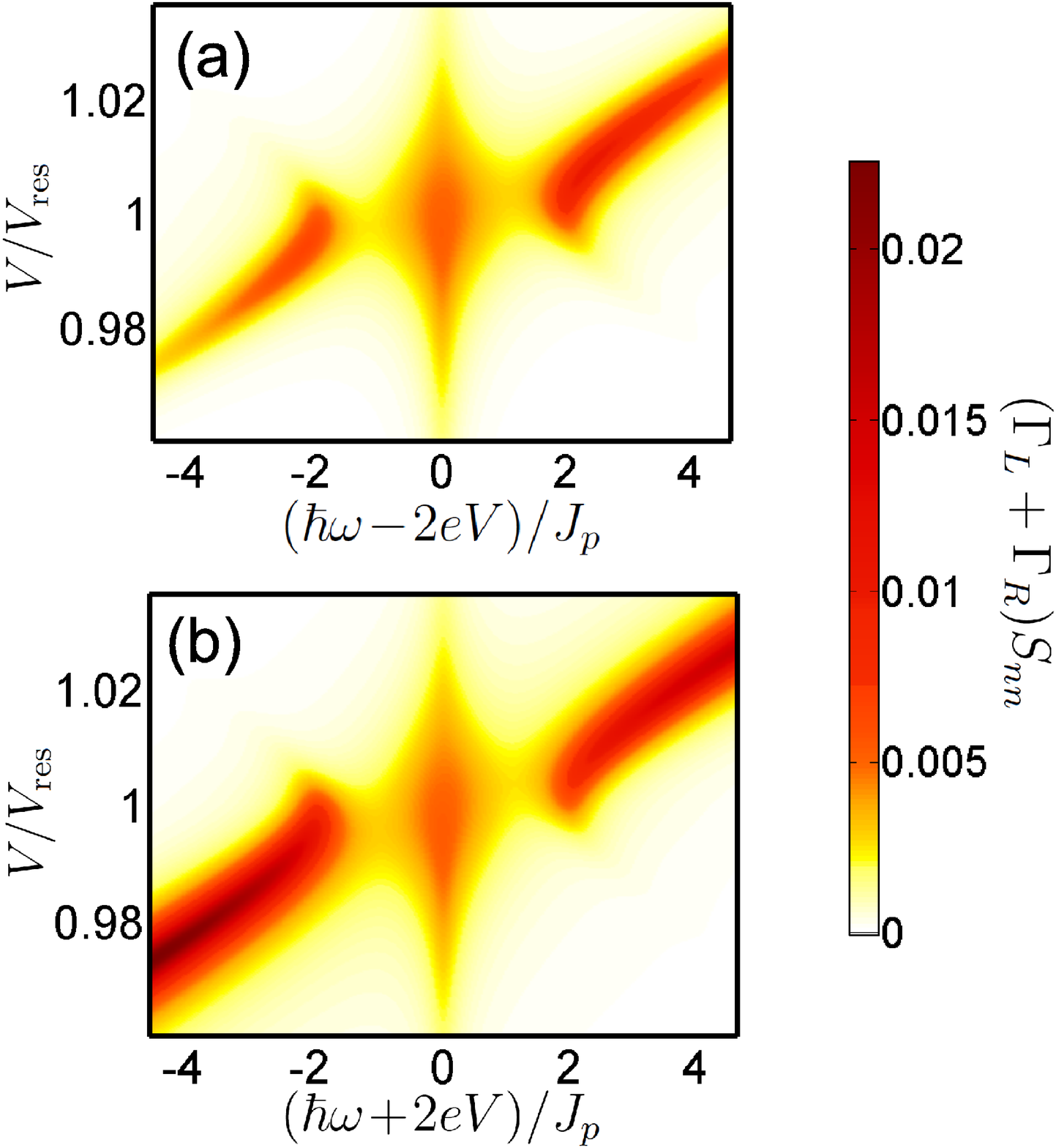, width=0.4\textwidth}
          \epsfig{file=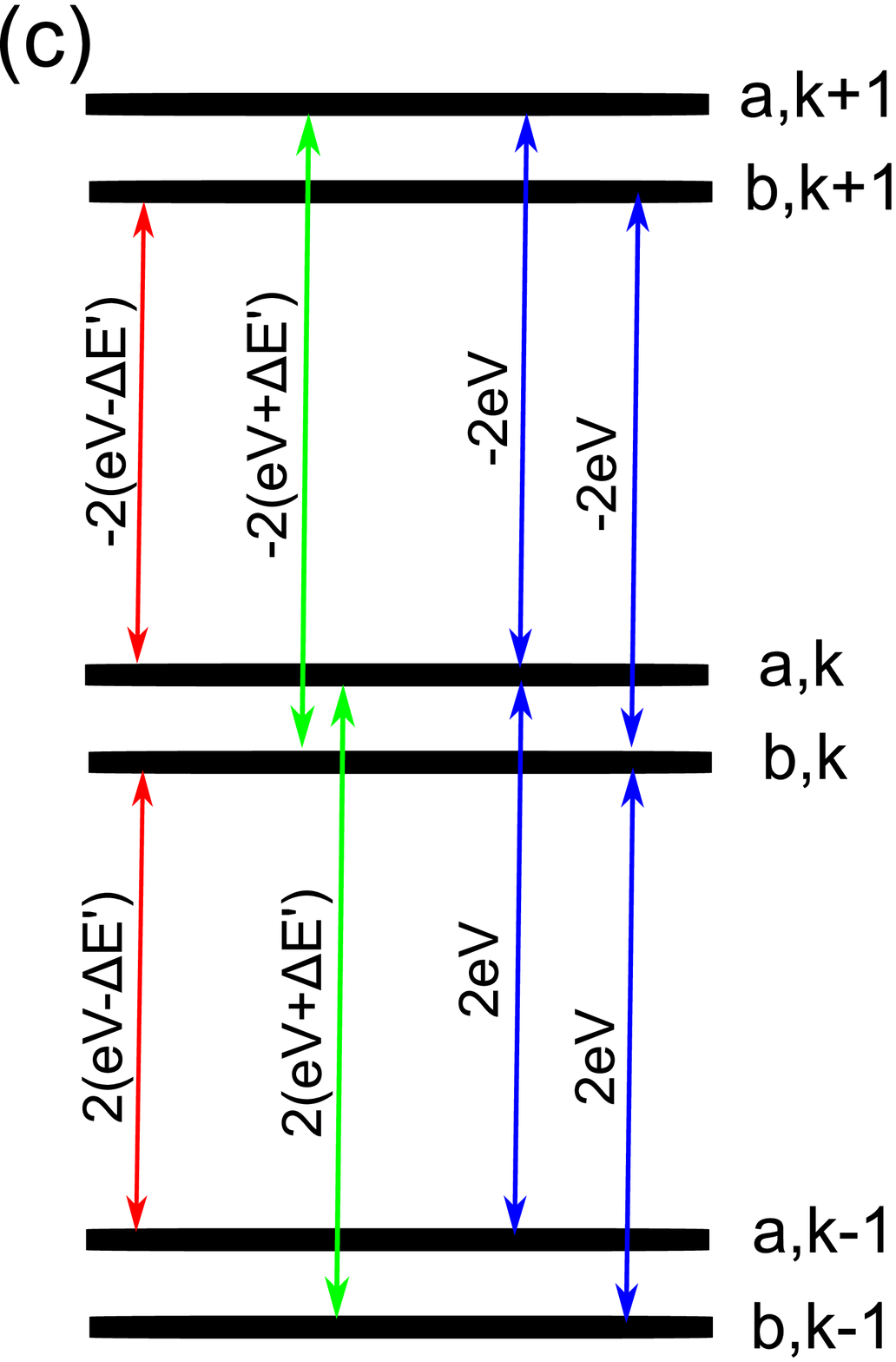, width=0.3\textwidth}
 \caption{(Color online) $S_{n^{(1)}n^{(1)}}(\omega)$ in the vicinity of the $p=1$ resonance (a) and (b) show the spectrum at frequencies around $\hbar\omega=\pm2eV$ respectively.  A schematic illustration of the processes giving rise to the spectral features is shown in (c). This illustrates the inter-doublet transitions and the corresponding frequencies of the features they give rise to.}\label{fig:n1n1}
\end{figure*}

The full charge noise spectrum also contains contributions at other frequencies far from $\omega=0$. These arise from the higher order contributions in the perturbative expansion of $n'$. In this section we investigate the most significant of these high order terms,  which arises from  the $\average{\nn{1}(t)\nn{1}(0)}$ term in the expansion. The form of $\nn{1}$ is calculated in the appendix,
\begin{multline}
  \nn{1}=-\sum_k \frac{J}{2peV}\ket{0,k}\bra{1,k +\half}\\+\frac{J}{2(p+1)eV}\ket{0,k}\bra{1,k-\half} +h.c.,
\end{multline}
for $p\geq 1$, in the case $p=0$ the first term is not present.  When this is transformed to the eigenstate basis it contains terms proportional
to $\ket{i,k}\bra{j,k+q}$, with $q=\pm p, \pm (p+1)$, for $p=0$ only terms $q=\pm1$ are present. As an example we calculate the corresponding spectrum,
\begin{equation}
  S_{\nn{1}\nn{1}}(\omega)=\int_{-\infty}^{\infty}\average{\nn{1}(t)\nn{1}(0)}{\rm e}^{i\omega t}dt,
\end{equation}
for the $p=1$ resonance (note that in the steady state  $\average{\nn{1}}=0$ and so we do not need to
work with the operator $\delta\nn{1}$). To do this we use the regression theorem and the reduced master
equations [Eq.\ \eqref{eqn:TLS}] with the values of $q=\pm 1, \pm 2$.

We find that the spectrum consists of triplets of peaks centered on the frequencies $\hbar\omega=2qeV$ with the sidepeaks separated
by the intra-doublet level spacing in each case. The triplets with $q=\pm 1$ and $q=\pm 2$ simply differ by a constant prefactor and a
slight modification to the decay rates and so we concentrate here on just the $q=\pm 1$ case. The spectrum around $\hbar\omega=\pm2eV$ is
shown in Figs.\ \ref{fig:n1n1}(a) and \ref{fig:n1n1}(b). The peaks in this part of the spectrum arise from first order Josephson coupling between states 
at the relevant frequency differences. We show the relevant transitions and their frequencies in Fig.\ \ref{fig:n1n1}(c). Since this spectrum occurs 
at higher order in the perturbation theory the magnitude of $S_{\nn{1}\nn{1}}$ is much smaller than it was for the zeroth order spectrum 
(Fig.\ \ref{fig:CNomega}). The central peak in both Figs.\ \ref{fig:n1n1}(a) and \ref{fig:n1n1}(b) is the same because the transitions 
which give rise to these peaks always link corresponding states within the two doublets (i.e.\ an $a$ state with an $a$ state or a $b$ state with a $b$) for both the positive and negative frequency processes as shown in Fig.\ \ref{fig:n1n1}(c). However, the sidepeaks of the two triplets are quite different. The weights of the sidepeaks are proportional to $|n_{ij}^q|^2\steady{ii}$ where $n_{ij}^q=\sum_k\bra{i,k+q}\nn{1}\ket{j,k}$ and $i(j)$ is the initial (final) state for the relevant transition. This means that each of the four sidepeaks has a different combination of matrix element and population. 

Far from the resonance only one of the three peaks is present in each triplet. In this region the eigenstates are very close to the charge states and since the Josephson effect only couples the states $\ket{0,k}$ and $\ket{1,k+\half}$  we find that each state is only coupled to one other.  Very close to the resonance all of the peaks appear, the eigenstates are mixtures of charge states and so transitions between all of the states in the doublets can occur.

The features seen in this part of the spectrum arise because the system is not as simple as a true TLS, they arise from
couplings between different doublets and hence require more than two energy levels. The frequencies at which the features in part of the
spectrum occur, $\sim100$\,GHz, are much larger than the range that can be probed with a stripline resonator. However, it might be possible to observe the noise at this frequency in a different kind of experiment in which the SSET is instead coupled to another mesoscopic conductor such as a SIS junction.\cite{billangeon:07, naaman:07}

\section{Conclusions}
\label{sec:conclusions}

We have analyzed the quantum dynamics of the SSET tuned close to Cooper-pair resonances. An effective Hamiltonian for the SSET was derived. this exploits the separation of the energy levels into doublets and accounts for the Josephson coupling between resonant states. We then derived the master equations for the system including the effect of the electromagnetic environment using the Born-Markov approximations. We calculated the current in the vicinity of the resonances and find, in accord with previous studies, strong peaks at the resonances.

 Calculating the charge noise spectrum in the vicinity of a resonance, we found that the spectrum is dominated by a triplet of peaks centered on zero frequency: a structure typical of a driven, damped, TLS.  However, the detailed form of the triplet in this case differed in important respects from a standard classically driven TLS because of the intra-doublet transitions. An experimentally realizable method of measuring the quantum noise spectrum is to couple the SSET to a resonator and measure the back-action. Using a linear response approach we found that the effects of the intra-doublet relaxation can be observed in the asymmetry of the damping rate, $\gba$, and effective occupation, $\nba$, of the resonator. It would also be possible to cool a low frequency oscillator using a Cooper-pair
 resonance, though not to the ground state as the large separation of the sidepeaks in the noise spectrum, $2\Jp/\hbar$, strongly limits the
 minimum occupation number that could be achieved. We also carried out a detailed calculation of the higher order triplet features that arise
  in the charge noise spectrum. These features occur because of the Josephson
 coupling between different doublets.

\section*{Acknowledgements}

We thank Miles Blencowe for useful discussions. We are grateful to the British Council (U.K.) and the French Foreign Affairs Ministry (France) for financial support through the program PHC Alliance.
Additional support from the EPSRC (U.K.) under grant EP/C540182/1 and ANR (France) through contract JCJC-036 NEMESIS is also acknowledged.

\appendix

\section{Transformations}
\label{app:HamTrans}
\subsection{Transforming the Hamiltonian}

We begin with the system Hamiltonian $H_S=H_{\text{ch}}+H_J$ as defined in Eqs.\ \eqref{eqn:Hch} and \eqref{eqn:HJ}. We find a unitary transformation, $H'=UHU^\dagger$ (we drop the subscript $S$ in this appendix), such that $H'$ only contains diagonal elements and those which couple resonant states. To do this we define the projector onto the $k$-th doublet as
\begin{equation}
  P_k=\ket{0, k}\bra{0, k}+\ket{1, k+p+\half}\bra{1, k+p+\half},
\end{equation}
then the condition on $H'$ above becomes $P_kH'P_{k''}=0$ for $k\neq k''$. We then write the transformation as\cite{cohen:92} $U={\rm e}^{iS}$ with $S=S^\dagger$ and treat $H_J$ as a perturbation. This allows us to write a series expansion for $S$ in terms of $J$,
\begin{equation}
  S=\Sn{0}+J\Sn{1}+J^2\Sn{2}+\ldots,
\end{equation}
 There are infinitely many transformations which satisfy the condition on $H'$ and so to uniquely specify $U$ we choose that $S$ should not have matrix elements within the doublet, $P_kSP_k=0$. We note that $\Sn{0}=0$, since for the case $J=0$ we need no transformation. This then allows us to find the transformed form of the Hamiltonian as
\begin{multline}
  H'=H_{\text{ch}}+H_J+[iJ\Sn{1}, H_{\text{ch}}]+[iJ\Sn{1}, H_J] \\ +[iJ^2 \Sn{2}, H_{\text{ch}}]+\frac{1}{2!} \left[ iJ\Sn{1}, [iJ\Sn{1}, H_{\text{ch}}]\right]+\ldots,
\end{multline}
which we can then write as a power series
\begin{equation}
  H'=\Hn{0}+\Hn{1}+\Hn{2}+\ldots,
\end{equation}
where we have grouped terms by order in $J$;
\begin{subequations}
\begin{gather}
  \Hn{0}=H_{\text{ch}}, \\
  \Hn{1}=H_J+[iJ\Sn{1}, H_{\text{ch}}], \\
\begin{split}
  \Hn{2}=[iJ\Sn{1}, H_J]+[iJ^2 \Sn{2}, H_{\text{ch}}]\\+\frac{1}{2!} \left[ iJ\Sn{1}, [iJ\Sn{1}, H_{\text{ch}}]\right].
\end{split}
\end{gather}
\end{subequations}
These can then be used to construct the expression for $S$ term by term and so build up the effective Hamiltonian. As an example we give the calculation of $\Sn{1}$. We begin by noting that $P_k\Hn{n}P_{k''}=0$ for all $n$ and $k\neq k''$ which gives,
\begin{equation}
  0=P_kH_JP_{k''}+P_k(iJ\Sn{1}H_\text{ch}-H_\text{ch}iJ\Sn{1})P_{k''},
\end{equation}
from which we can calculate the matrix elements of $i\Sn{1}$ as
\begin{gather}
  \bra{0,k}i\Sn{1}\ket{1,k+\half}=-\frac{J}{2peV}=G_1, \\
  \bra{0,k}i\Sn{1}\ket{1,k-\half}=-\frac{J}{2(p+1)eV}=-G_2,
\end{gather}
and the obvious conjugates. $G_1$ and $G_2$ then become the natural small parameters of the perturbation theory. These expressions can then be used to calculate $\Sn{2}$ and so on. This allows the transformed Hamiltonian to be found up to any order, for example we find that the second order corrections to the diagonal elements are given by
\begin{equation}
  \bra{0,k}\Hn{2}\ket{0,k}=-\frac{J^2}{eV}\frac{2q}{q^2-1}.
\end{equation}
We also need to be able to calculate the off-diagonal coupling between resonant states, $J_p$. This appears to order $2p+1$ in the perturbation expansion. The method outlined above becomes very cumbersome at high orders and so to calculate $J_p$ for higher orders we introduce the level-shift operator\cite{cohen:92, joyez:95}
\begin{multline}
R(z)=H_J+H_J\frac{Q_k}{z-H_{\text{ch}}}H_J\\+H_J\frac{Q_k}{z-H_{\text{ch}}}H_J\frac{Q_k}{z-H_{\text{ch}}}H_J+\ldots,
\end{multline}
where $Q_k=1-P_k$. Which then allows us to calculate $\Jp=\bra{0,k}R(\bar{E})\ket{1,k+p+\half}$. This gives the expression found in Eq.\ \eqref{eqn:jp}. However, we have checked explicitly that this approach gives the same expression for $J_{p=1}$ (the highest order used in the main text) as that obtained using the perturbation expansion.

\subsection{The transformed operators}
We also need to transform the operators $k$ and $n$. To illustrate the method we give an explicit calculation of $k'$. This is done in exact analogy with the calculation of $H'$. We first write $k'$ as a power series,
\begin{equation}
  k'=\kn{0}+\kn{1}+\kn{2}+\ldots,
\end{equation}
where,
\begin{subequations}
\begin{gather}
  \kn{0}=k, \\
  \kn{1}=[iJ\Sn{1}, k], \\
  \kn{2}=[iJ^2\Sn{2}, k]+\frac{1}{2!} \left[ iJ\Sn{1}, [iJ\Sn{1}, k]\right],
\end{gather}
\end{subequations}
which then allow us to calculate $\kn{1}$ as
\begin{multline}
  \kn{1}=\frac{1}{2}\sum_k G_1\ket{0,k}\bra{1,k+\half}\\ + G_2\ket{0,k}\bra{1,k-\half} +h.c., \label{eqn:kn1}
\end{multline}
and $\kn{2}$ is given by
\begin{widetext}
\begin{multline}
  \kn{2}=\frac{(2p+1)G_1G_2}{2}\sum_k \left( \ket{0,k}\bra{0,k+1}-\ket{0,k}\bra{0,k-1} + h.c.\right) \\
                 +(G_1^2+G_2^2)\sum_k(\ket{1,k+\half}\bra{1,k+\half}-\ket{0,k}\bra{0,k}).
\end{multline}
\end{widetext}
A similar calculation can be performed to calculate the power series for $n'=n^{(0)}+n^{(1)}+n^{(2)}+\ldots$, and we find $n^{0}=n$,
\begin{equation}
  \nn{1}=\sum_k G_1\ket{0,k}\bra{1,k+\half} - G_2\ket{0,k}\bra{1,k-\half} +h.c.,
\end{equation}
\begin{multline}
  \nn{2}=G_1G_2\sum_k \left( \ket{0,k}\bra{0,k+1}-\ket{0,k}\bra{0,k-1} + h.c\right) \\
                 +(G_1^2+G_2^2)\sum_k(\ket{1,k+\half}\bra{1,k+\half}-\ket{0,k}\bra{0,k}).
\end{multline}


\begin{thebibliography}{99}
\bibitem{footnote3} We evaluate $\Jp$ and the second order correction to $\Delta E$ using the resonance value for the voltage, $V=V_{\rm res}^{(p)}$, as these terms only give a significant contribution in the vicinity of the resonances when the electrostatic part of $\Delta E$ is small. Making this approximation simplifies the calculation of the perturbation series.
\bibitem{footnote} Secular terms are those for which $|\omega_{\mu\nu}-\omega_{\mu'\nu'}|\ll1/\Delta t$ where $\Delta t$ is the coarse-graining timescale 
(much larger than the correlation time of the environment, but much shorter than the timescale over which $\tilde{\sigma}$ evolves) over 
which the master equations are valid. Only terms where $\omega_{\mu\nu}=\omega_{\mu'\nu'}$ are included
since the smallest Bohr frequency in the system is $\Jp/\hbar$ which we can safely assume is much larger than $1/\Delta t$,
\bibitem{footnote2} In our calculations of the current and charge noise in the vicinity of the resonances we
generally evaluate $\Gamma_{L,R}$  using the resonance value for the voltage, $V=V^{(p)}_{\rm res}$, as within the range of voltages over which the doublet picture is valid these quantities are typically very slowly varying.
\end{thebibliography}
\end{document}